\documentclass[aps,superscriptaddress,preprint,nofootinbib,eqsecnum]{revtex4-1}
\usepackage{graphicx,amsmath,latexsym}
\newcommand{\beq}{\begin{equation}}
\newcommand{\eeq}{\end{equation}}
\newcommand{\bqa}{\begin{eqnarray}}
\newcommand{\eqa}{\end{eqnarray}}
\newcommand{\f}{\frac}
\newcommand{\dr}{\partial_r}
\newcommand{\nn}{\nonumber}
\newcommand{\rd}{\right.}
\newcommand{\ld}{\left.}

\begin{document}

\title{Shock wave collisions in $AdS_5$: approximate numerical solutions}

\author{Bin Wu}
\affiliation{
Frankfurt Institute for Advanced Studies,
    D-60438 Frankfurt, Germany
}
\affiliation{Faculty of Physics, University of Bielefeld, D-33501 Bielefeld, Germany}
\affiliation{School of Physics and State Key Laboratory of Nuclear Physics and Technology, Peking University, Beijing 100871, China}
\author{Paul Romatschke}
\affiliation{
Frankfurt Institute for Advanced Studies,
    D-60438 Frankfurt, Germany
}
\affiliation{Department of Physics, 390 UCB, University of Colorado, Boulder,
CO 80309, USA}

\begin{abstract}
\noindent We numerically study the evolution of a boost-invariant  ${\cal N}=4$ SYM medium using AdS/CFT. 
We consider a toy model for the collision of gravitational shock waves,
finding that the energy density first increases, reaches a maximum 
and then starts to decrease, matching hydrodynamics for late times.
For the initial conditions we consider, 
the hydrodynamic scale governing the late time behaviour is to very good approximation determined by the area of the black hole
horizon at initial times. Our results provide a toy model for the 
early time evolution of the bulk system in heavy-ion collisions at RHIC and the LHC.
\end{abstract}


\maketitle

\section{Introduction}

The problem of colliding gravitational shock waves in spaces that are 
asymptotically Anti-de-Sitter has been of recent interest
because it can serve as a toy model of the collision
of two nuclei approaching at very high speeds. Hence it
may provide --- via the AdS/CFT conjecture\cite{Maldacena,Witten} --- 
some qualitative insight in phenomena found
in heavy-ion experiments at the Relativistic Heavy-Ion
Collider (RHIC) and the Large Hadron Collider (LHC).

Boosting a mass at rest to very high velocities, its energy-momentum
tensor in coordinates $x^{\pm}=\frac{t\pm z}{\sqrt{2}}$ becomes
that of a gravitational shock wave, e.g. $T_{++}\propto \mu \delta(x^+)$,
where $\mu$ is the energy per unit area which for a nucleus of
atomic number $a$, radius $R$ and Lorentz boost factor 
$\gamma=\frac{\sqrt{s_{NN}}}{2m_p}$
is 
\beq
\label{mudef}
\mu\propto \gamma \frac{a m_p}{\pi R^2}= \frac{a \sqrt{s_{NN}}}{2\pi R^2}\,,
\eeq
where $m_p$ is the proton mass. It is well known how to model 
such an energy-momentum tensor using the AdS/CFT correspondence,
namely by a metric of the form \cite{Grumiller:2008va} 
(cf.\cite{AlvarezGaume:2008fx,Lin:2009pn,Albacete:2009ji})
\beq
\label{singlelineE}
ds^2=\frac{-2 dx^+ dx^- + \phi(z)\delta(x^+) dx^{+2}+dx_{\perp}^2+dz^2}{z^2}
\eeq
where $\phi(z)=\frac{\mu z^4}{\kappa}$ and we have set
the $AdS$ radius to unity. Here the description
assumes the conjectured duality between ${\cal N}=4$ SYM at
 large coupling and large number of colours $N_c$ and classical
gravity on $AdS_5\times S^5$. Since ${\cal N}=4$ SYM is a 
gauge theory, it behaves qualitatively similar to QCD, 
so some aspects of this work may translate to qualitatively
similar phenomena found in nature. The constant $\kappa$ is usually
set to $\frac{N_c^2}{2 \pi^2}$, but we will treat $\kappa$ as a free
parameter to be adjusted at will in order to obtain a model
that more closely resembles QCD.

A collision of two nuclei can be modelled by a superposition
of two shock waves in $AdS_5$, moving in $x^+$ and $x^-$ direction,
respectively. While the line element before the collision
is a simple superposition of the individual shock waves,
the metric after the collision is in general hard to find. 
Unfortunately, while exact analytic solutions in four dimensional
asymptotic Minkowski time have been derived many years
ago~\cite{Khan:1971vh}, no such solutions are known for shock
waves of the form (\ref{singlelineE}). Therefore,
one has to resort to numerical techniques, which were
pioneered in \cite{cy1,cy2}, see also \cite{Chesler:2010,Heller:2011ju}.

Another aspect of Eq.~(\ref{singlelineE}) is
that the collision of two such shock waves can be shown
to violate Bjorken's conjectured invariance under 
rapidity boosts (\cite{Grumiller:2008va}). Since experimental
data for heavy-ion collisions 
does not seem to back up this invariance either, this can be considered
a feature rather than a shortcoming of the present model,
but at the price that the gravitational dynamics is 2+1 dimensional
(rapidity, AdS radius and time)~\cite{Chesler:2010} rather than 1+1 dimensional.

However, as we shall point out in the present work, it turns
out that in the limit of weak shock waves $\mu\ll1$,
the leading order dynamics is in fact boost-invariant
(cf. Ref.~\cite{Bhattacharyya:2009uu}).
The late-time behavior of such a strongly coupled boost-invariant ${\cal N} = 4$ SYM medium 
has been known up to the $3^{rd}$ order in large $\tau$ expansion~\cite{Janik:2005,Janik:2006,Kinoshita:2008dq, Booth:2009ct}. However, one has to use numerical methods to fully understand the early-time properties of the system~\cite{Beuf:2009cx}. In this paper we use 
algorithms similar to those in Refs. \cite{cy1,cy2} to solve Einstein's equations numerically 
in this approximation, and follow the evolution of the 
boundary energy-momentum tensor from the far-from equilibrium situation
at early times to the hydrodynamic behavior at late times. 
Unlike Ref.~\cite{cy1,cy2}, we do not deform the boundary four dimensional metric of the $AdS_5$ space but connect initial conditions
derived analytically from the shock waves before the collision
to the late time hydrodynamic regime. Our findings validate those of Ref.~\cite{Heller:2011ju}, where the 
authors use a different algorithm and start with arbitrary initial conditions.

This paper is organized as follows. In Sec.~\ref{sec:shock} we construct an ansatz metric function based on the approximate metric functions in the collisions of two weak shock waves. In Sec.~\ref{sec:numerics} two algorithms for numerically solving Einstein's equations in the bulk of the $AdS_5$ space are described in detail. Our numerical results and an application to RHIC and LHC are
presented in Secs.~\ref{sec:results},\ref{sec:early}.
In the Appendix \ref{app:A} provide near-boundary 
power series expansions needed in our numerical calculations.

\section{Collisions of two weak shock waves}\label{sec:shock}

The line element (\ref{singlelineE}) is highly singular at $x^\pm=0$, and it 
is useful to first change to so-called Rosen coordinates
$x^+=u\,,\quad x^-=v+\frac{1}{2} \phi(\tilde{z}) \theta(u)+\frac{1}{8} \left(\phi'(\tilde{z})\right)^2 u \theta(u)^2\,,\quad
z=\tilde{z}+\frac{1}{2}\phi'(\tilde{z}) u \theta(u)\,,$
with the result
\bqa
\label{singleshock}
&ds^2=\frac{-2 du dv+dx_\perp^2+\left(1+\frac{1}{2}\phi''(\tilde{z}) u \theta(u)\right)^2d\tilde{z}^2}{\left(\tilde{z}+\frac{1}{2}\phi'(\tilde{z})u \theta(u)\right)^2}\,.&
\end{eqnarray}

The collision of two shocks can be set up by superposing the above 
line element for one shock with an equivalent one for the other shock.
The difficult part of the calculation then involves finding the line
element in the forward light-cone. Using the standard matching
conditions (metric needs to be continuous and piece-wise differentiable
\cite{Khan:1971vh}) it has been possible to find the metric 
in the approximation of small strength $\mu$
\cite{Grumiller:2008va}. Using the coordinates proper time 
$\tilde \tau=\sqrt{2 u v}$ and space-time rapidity $\tilde \eta=\frac{1}{2}\ln\frac{u}{v}$ 
the result is given by \cite{Grumiller:2008va}
\bqa
ds^2&=&\frac{-\left[1+K(\tilde\tau,\tilde\eta,\tilde{z})\right]d\tilde\tau^2
+\left[1+L(\tilde\tau,\tilde\eta,\tilde{z})\right]\tilde\tau^2 d\tilde\eta^2 
+\left[1+H(\tilde\tau,\tilde\eta,\tilde{z})\right]d{\bf x_\perp}^2}{\tilde{z}^2
\left[1
+2 \tilde{z}^2 \bar\mu \tilde\tau \cosh(Y-\tilde\eta)
\right]^2}\nonumber\\
&&\hspace*{4cm}
+\frac{\left[1+M(\tilde\tau,\tilde\eta,\tilde{z})\right]
\left[1+6 \tilde{z}^2 \bar\mu \tilde\tau \cosh(Y-\tilde\eta)
\right]^2 d\tilde{z}^2}{\tilde{z}^2
\left[1+2 \tilde{z}^2 \bar\mu \tilde\tau\cosh(Y-\tilde\eta)\right]^2}\,,
\label{oursol}
\eqa
where $\bar\mu=\mu/\kappa$ and $K,L,H,M$ were determined to be 
\bqa
K(\tilde\tau,\tilde\eta,\tilde{z})&=&c_1 \bar\mu^2 \tilde\tau^2 \tilde{z}^4
-\frac{5+c_1}{3}\bar\mu^2 \tilde\tau^4 \tilde{z}^2 +{\cal O}(\bar\mu^3)
\nonumber\\
L(\tilde\tau,\tilde\eta,\tilde{z})&=&\frac{-16+c_1}{3} \bar\mu^2 \tilde\tau^2 \tilde{z}^4
-\frac{5+c_1}{3}\bar\mu^2 \tilde\tau^4 \tilde{z}^2
+{\cal O}(\bar\mu^3)
\nonumber\\
H(\tilde\tau,\tilde\eta,\tilde{z})&=&-2 \bar\mu^2 \tilde\tau^2 \tilde{z}^4
-\frac{5+c_1}{3}\bar\mu^2 \tilde\tau^4 \tilde{z}^2+{\cal O}(\bar\mu^3)\nonumber\\
M(\tilde\tau,\tilde\eta,\tilde{z})&=&16 \bar\mu^2 \tilde\tau^2 \tilde{z}^4
+\frac{10+2 c_1}{3}\bar\mu^2 \tilde\tau^4 \tilde{z}^2+{\cal O}(\bar\mu^3)
\label{solfunc}
\eqa
by solving Einstein's equations $R_{\mu \nu}-\frac{1}{2}g_{\mu \nu} R-6 g_{\mu \nu}=0$. Here $c_1$ is a freely choosable integration constant that
corresponds to some unfixed diffeomorphism freedom.
Following Refs.\cite{cy1,cy2}, our numerical setup requires a line-element
in the Eddington-Finkelstein form and, therefore, we have to transform to new coordinates
$\tau,\eta,z$. The relation between the old and new coordinates in the 
small $\bar\mu$ limit may be found to be 
\bqa
\tilde \tau &=& \tau + z  -\frac{\bar\mu}{2}z^4 \cosh{[Y-\eta]}
-\frac{1}{2}\bar\mu^2 z^4 (\tau+z)^3 c_1
+\bar\mu^2 \cosh{[2 (Y-\eta)]} z^6 (\tau+z)\nonumber\\
&&-\frac{\bar\mu^2}{210}z^4 (175 \tau^3+469 \tau^2 z+378 \tau z^2+52 z^3)
-\frac{\bar\mu^2 z^8 \sinh^2{[Y-\eta]}}{8 (\tau+z)}+{\cal O}(\bar\mu^3)\nonumber\\
\tilde \eta &=&\eta - \frac{\bar\mu z^4 \sinh{[Y-\eta]}}{2 (\tau+z)}
+\frac{\bar\mu^2 (8 \tau^2 +16 z +7 z^2)}{8 (\tau+z)^2} z^6 \sinh{[2 (Y-\eta)]}
+{\cal O}(\bar\mu^3)\nonumber\\
\tilde z &=& z -2 \bar\mu z^3 (\tau+z) \cosh{[Y-\eta]}-\frac{\bar\mu^2 z^3 (\tau+z)^4 c_1}{6}+\bar\mu^2 \cosh{[2 (Y-\eta)]} 6 z^5 (\tau+z)^2\nonumber\\
&&-\frac{\bar\mu^2}{30} z^3
\left(25 \tau^4+100 \tau^3 z+25 \tau^2 z^2-146 \tau z^3-119 z^4\right)
+{\cal O}(\bar\mu^3)\,.
\eqa

It turns out that to this order in the shock strength $\mu$, the metric 
in Eddington-Finkelstein coordinates is independent of $\eta$
and hence boost-invariant in the sense of Bjorken~\cite{Bjorken:1982qr}. Higher order corrections
turn out to spoil this invariance, but it seems that --- at least for
weak shocks with $\mu\ll1$ --- the initial dynamics is predominantly
boost-invariant. In Eddington-Finkelstein coordinates, the line element can be parametrized in the following form
\beq
\label{CYlineelement}
ds^2=2 dr d\tau - A d\tau^2 + \Sigma^2 e^B dx_\perp^2+\Sigma^2 e^{-2 B} d\eta^2
\,,
\eeq
where using $r=\frac{1}{z}$ the metric functions are given as
\bqa
\label{perturbative}
A &=&r^2-\f{6 \bar \mu^2}{5 r^4}  - \f{4 \bar \mu^2 \tau}{3 r^3} - 
\f{4 \bar\mu^2 \tau^2}{3 r^2}+{\cal O}(\bar \mu^3)\,,\nonumber\\
B&=&-\frac{2}{3} {\log}\left(\frac{1+r \tau}{r}\right)+\frac{(612+7 r \tau (234+r \tau (217+75 r \tau))) {\bar\mu}^2}{315 r^6 (1+r \tau)}+{\cal O}(\bar \mu^3)\,,\nonumber\\
\Sigma^3&=&r^2 (1 + r \tau) + \f{(72 + 14 r \tau (9 + 5 r \tau)) \bar\mu^2}{105 r^4}+{\cal O}(\bar\mu^3)\,.\label{equ:metricshockapp}
\eqa

\subsection{Einstein's equations in Eddington-Finkelstein coordinates}

In Eddington-Finkelstein coordinates (\ref{CYlineelement}), Einstein's 
equations become \cite{cy1}
\begin{subequations}
\begin{eqnarray}
\label{equ:eea}
0 &=& \Sigma \,  \dot \Sigma{'} + 2 \Sigma{'} \, \dot \Sigma - 2 \Sigma^2\,,
\\ \label{equ:eeb}
0 &=& \Sigma \, \dot B{'} + {\textstyle \frac{3}{2}}
    \big ( \Sigma{'} \dot B +  B{'} \, \dot \Sigma \big )\,,
\\  \label{equ:eec}
0 &=& A{''} + 3  B{'} \dot B - 12 \Sigma{'} \, \dot \Sigma/\Sigma^2 + 4\,,
\\  \label{equ:eed}
0 &= & \ddot \Sigma
    + {\textstyle \frac{1}{2}} \big( \dot B^2 \, \Sigma -  A{'} \, \dot \Sigma \big)\,,
\\ \label{equ:eee}
0 &=& \Sigma{''} + {\textstyle \frac{1}{2}} B'^2 \, \Sigma\,,
\end{eqnarray}
\label{Eeqns}%
\end{subequations}
where for any function $h(r,\tau)$ we have defined
\begin{equation}
    \dot h \equiv \partial_\tau h + {\textstyle \frac{1}{2}} A \, \partial_r h\,,
\end{equation}
and $h'\equiv\dr h$.
Under the coordinate transformation
\beq
r \rightarrow \hat{r} = r - f,\label{equ:diffeomorphism}
\eeq
one has
\beq
\hat{A} = A( \hat{r}+ f, \tau ) - 2 \f{df}{d\tau} ,~\hat{\Sigma}=\Sigma( \hat{r}+f , \tau),~\mbox{and}~\hat{B}=B( \hat{r}+f, \tau ),\label{equ:metrichat}
\eeq
where $f$ is an arbitrary function of $\tau$. It is easy to check that Einstein's equations are form-invariant under the above diffeomorphism.

In the following, we will take (\ref{equ:eed}) and (\ref{equ:eee}) as constraint equations and numerically solve (\ref{equ:eea})-(\ref{equ:eec}), which can be rewritten in the following form
\begin{subequations}
\bqa
&& \theta' = S,\label{equ:theta}\\
&&\phi'= 3 B'  S^{-\f{1}{2}} \theta,\label{equ:phi}\\
&&\dot{S} = 6 \theta\label{equ:dotS},\\ 
&&\dot{B} = - S^{-\f{1}{2}} \phi\label{equ:dotB},\\
&&A'' = 8  \theta S' S^{-2} + 3 B' \phi S^{-\f{1}{2}} - 4,\label{equ:App}
\eqa
\end{subequations}
where $S \equiv \Sigma^3$.

\subsection{Apparent Horizons and Area of Trapped Surface}

The area of the trapped surface formed in the collision
of two shock waves is of considerable interest since
at late times, when the system is close to equilibrium, it
can be used to extract the entropy of the system. Far from
equilibrium, its physical interpretation is difficult \cite{Booth:2009ct}
but it is nevertheless interesting to track the area spanned
by the apparent horizon, 
which is the location where out-going null vectors vanish.
It should be pointed out that the apparent horizon is a local concept,
coordinate-time dependent, and not invariant under coordinate
transformations (space-time slicings).

The location of apparent horizon may be calculated as follows:
first determine the in and out-going null vectors $l^-,l^+$
(corresponding to ``light-rays'' in ordinary space-time) from the 
condition $g^{\mu\nu}l_\mu l_\nu=0$. Then find the apparent horizon
from the criterion of vanishing expansion of the out-going null vectors,
\begin{equation}
\label{expansion}
h^{ab} \nabla_a l_b^+=0\,,
\end{equation}
where $h^{ab}$ is the projected metric that is given by
$
h_{ab}=g_{ab}-\frac{l_a^+l_b^-+l_a^- l_b^+}{l^+ \cdot l^-}\,.
$ 
(The projected metric fulfills the requirement that 
multiplying $l^+$ by an arbitrary function $B$ does not change the 
result (\ref{expansion})).

\subsubsection{Before the collision}
\label{sec:before}

Just before the collision of the two shock waves, where the 
line element is given by a superpositions of line elements
of the form (\ref{singleshock}), the location 
of the apparent horizon may be calculated along the lines of 
\cite{Gubser:2008pc}:
parametrizing the surface at $u=0$ by $v=-\psi_1(\tilde{z})$, 
the normals to this surface are given by 
$du,dv+d\psi_1$, so normal vectors $l_\mu$ can be parametrized as 
$
l_\mu dx^\mu=c_1 du+c_2 (dv+d\psi_1)\,.
$
The condition $g^{\mu\nu}l_\mu l_\nu=0$ at $u=0$ leads to the conditions
$
c_1=c_2 \frac{\psi_1^{\prime 2}(\tilde{z})}{2}\quad {\rm or}\quad c_2=0\,,
$
where the prime here denotes a derivative with respect to $\tilde z$.
As a consequence we obtain a set of (out-going and in-going) null vectors
$l_\mu^{\pm}$ normal to the surface at $u=0$
$$
l_\mu^{+}=\left(\frac{\psi_1^{\prime 2}(\tilde{z})}{2},1,{\bf 0}_\perp,\psi_1'(\tilde z)\right)\,,\quad 
l_\mu^{-}=\left(1,0,{\bf 0}_\perp,0\right)\,.
$$
(The constant function multiplying these vectors is arbitrary and
has been set to unity.)
Vanishing expansion implies
$$
\Box \left[\psi_1(\tilde z)-\frac{1}{2}\phi(\tilde z)\right]=0\,,
$$
which has the solution $\psi_1(\tilde z)=\frac{1}{2}\phi(\tilde z)+c$,
where the constant $c$ is unimportant for the following.
To obtain the location of the trapped surface $\tilde z=\tilde z_H$,
we use the following boundary condition:
we could have equally well started with the other shock wave and
a surface at $v=0$ parametrized as $u=-\psi_2(\tilde z)$. Since
at $u=v=0$ both surface normal vectors have to coincide, one finds
$\psi_1=\psi_2$ and $\psi_1^{\prime 2}(\tilde z_H)=2=\frac{1}{4}\phi^{\prime 2}(\tilde z_H)$
and as a consequence
$$
\tilde z_H=\left(2 \bar \mu^2\right)^{-1/6}\,.
$$
Now the ``area'' of the trapped surface is given by
$$
A_h=\int \sqrt{{\rm det} \left.g_{ab}\right|_S} dx_\perp d\tilde z=
\int dx_\perp \int_{\tilde z_H}^{\infty}\frac{1}{\tilde z^3} d\tilde z
= \left(\frac{\bar\mu}{2}\right)^{2/3}\int dx_\perp
$$
where $\left.g_{ab}\right|_S$ is the induced metric on the trapped surface,
which can be calculated by using $du=0,dv+d\psi_1(\tilde{z})=0$ in
the line element. 

\subsubsection{After the collision for weak shocks}

For the line element (\ref{CYlineelement}) we parametrize the
location of the apparent horizon by $r=r_h(\tau)$ at constant $\tau$,
which leads to
$
l_\mu dx^\mu = \left(c_1 - c_2 r_h^\prime\right)d\tau + c_2 dr\,.
$
The conditions for null vectors are $c_1=-\frac{1}{2}c_2 A+c_2 r_h^\prime$
and $c_2=0$, so that for constant $\tau$
$$
l^+_\mu=\left(-\frac{1}{2} A,0,0,0,1\right)\,\qquad
l^-_\mu=\left(1,0,0,0,0\right)\,.
$$
Vanishing expansion implies 
$\dot \Sigma\equiv\partial_\tau \Sigma+\frac{1}{2}A \partial_r \Sigma=0$,
or equivalently $\dot S=0$ where we recall that $S\equiv \Sigma^3$.
Because of (\ref{equ:dotS}), an equivalent condition is 
$\theta(r=r_h)=0\,,$
which
is sometimes easier to use because the definition of $\theta$ does neither involve
the function $A$ nor explicit time derivatives.

Using the values for $A,\Sigma$ from (\ref{perturbative}) to solve 
$\dot S(r=r_h)={\cal O}(\mu^3)$ (or equivalently 
integrating $S$ to obtain $\theta(r)$ and solving for $\theta(r=r_h)={\cal O}(\mu^3)$) one finds
in the limit $\tau\rightarrow 0$ 
\beq
\label{perthor}
r_h(\tau=0)\sim\left(\frac{24 \bar\mu^2}{35}\right)^{1/6}
\eeq
and the horizon area becomes
\beq
\label{pertarea}
A_h(\tau=0)=\int dx_\perp d\eta \Sigma^3\sim\frac{192^{1/3} \bar\mu^{2/3}}{35^{1/3}}V\,,
\eeq
with $V=\int dx_\perp d\eta$.

This result is qualitatively the same as in Sec. \ref{sec:before},
which is encouraging. However, there may be sizeable quantitative
corrections to the above numbers, which can be 
traced back to the approximation used
in deriving (\ref{perturbative}), namely small $\mu$.
In terms of
the variable $z=1/r$ it is apparent that while
corrections ${\cal O}(\mu^3)$ to (\ref{perturbative}) are suppressed
close to the boundary $z=0$, they become of order unity when 
$\bar\mu z^3\sim 1$, or $r\sim \bar\mu^{1/3}$. Hence, the line
element is not a valid approximation close
to the apparent horizon and in particular will not fulfill
Einstein's equations there. For this reason, we will make 
an ansatz for the line element in Eddington-Finkelstein coordinates
that corresponds to (\ref{perturbative}) for small $\mu$, but
is a solution to Einstein's equations everywhere.

\subsection{An ansatz for the post-collision line element}\label{sec:metricansatz}

The ansatz we choose for the line element is to take 
\beq
\label{sigmaansatz}
\Sigma^3=r^2 (1 + r \tau) + \f{(72 + 14 r \tau (9 + 5 r \tau)) \bar\mu^2}{105 ( r^4 + \bar{c} \bar \mu^{\f{4}{3}}) },
\eeq
where $\bar{c}$ is a positive constant ('fudge parameter'). This ansatz agrees with the
``perturbative'' result (\ref{perturbative}) in the limit
of small $\bar\mu$ and/or small $\bar c$.
In the scheme we will employ, 
the coefficient functions $A,B$ can be calculated numerically
from Einstein's equations. We are then able to study a toy model
for shock wave collisions that involves one unknown number, $\bar c$.

On the CFT side, the above initial geometry corresponds to a 
strongly coupled gauge theory (${\cal N}=4$ SYM) with the energy density
\beq
\label{ED}
\epsilon\equiv T_{\tau\tau}=\kappa \bar\mu^2 \tau_0^2+{\cal O}(\tau_0^3)\,.
\eeq
This initial condition implies that the initial energy density
does not depend on the choice for $\bar c$ (nor does 
any other component of the CFT stress tensor). 

By contrast, the area of the apparent horizon $A_h(\tau)=S(r_h, \tau)$
depends significantly on $\bar c$. 
For the gauge choice $f=0$ (see (\ref{equ:diffeomorphism})) \
and in the limit $\tau_0\rightarrow 0$ and $\bar c\ll1$, 
the horizon position and area correspond to Eqs.~(\ref{perthor}) and (\ref{pertarea}), 
respectively, while for $\bar c\gg 1$ they are given by 
$r_h(\tau_0\ll 1,\bar c\gg 1)=\left(\frac{648 \pi^2\bar \mu^2}{1225 \bar c^{3/2}}\right)^{1/6}$
and $A_h(\tau_0\ll 1,\bar c\gg 1)=r_h^2 V$.
Since $A_h$ is a monotonously decreasing function of $\bar c$, we may try to approximate
its behaviour by the Pad\'e inspired ansatz
\beq
\label{PDinitialA}
A_h(\tau_0\ll 1)=\bar \mu^{2/3} V \frac{k_0}{\sqrt{k_1+\bar{c}}}\,,
\eeq
fixing the constants $k_0,k_1$ from the known large and small $\bar c$ limits as
$k_0=\left(\frac{648 \pi^2}{1225}\right)^{1/3}$, $k_1=\left(\frac{27 \pi^2}{280}\right)^{2/3}$.
We find that this ansatz gives a quite accurate approximation of the numerically determined
horizon area for arbitrary $\bar c$ in the limit of small $\tau_0$.

Unfortunately, not all values of $\bar c$ lead to physically
acceptable initial conditions, because (\ref{equ:eee}) 
requires that $3 S \partial_r^2 S-2 (\partial_r S)^2<0$ for
$r>r_h$, which is not fulfilled for any $\bar c$. Specifically,
$3 S \partial_r^2 S-2 (\partial_r S)^2>0$ for $r>r_V$, with
$r_V(\tau_0,\bar c)$ a monotonously increasing function of $\bar c$.
The condition $r_V(\tau_0,\bar c)<r_h(\tau_0,\bar c)$ leads to 
the requirement $\bar c(\tau_0)>\bar c_{\rm min}(\tau_0)$, with
$c_{\rm min}$ specified in Tab.~\ref{tab:smallestcbar}.

\begin{table}
\caption{Smallest allowed $\bar c$ at different initial times.}\label{tab:smallestcbar}\hspace{2mm}
 \centerline{
\begin{tabular}{c||ccccccc}
\hline\hline
$\tau_0$&0&0.1&0.2&0.3&0.4&0.5&1.0\\\hline
~~$\bar c_{\rm min}(\tau_0)~~$&~~2.88~~&~~3.43~~&~~4.10~~&~~4.93~~&~~5.97~~&~~7.26~~&~~19.84~~\\
~~$V^{-1}\bar\mu^{-2/3}A_h(\tau_0,\bar c_{\rm min})$~~&~~0.83~~&~~0.92~~&~~1.02~~&~~1.15~~&~~1.30~~&~~1.47~~&~~2.85~~\\
\hline\hline
\end{tabular}}
\end{table}

\section{Numerics}\label{sec:numerics}

Near the boundary $r\rightarrow \infty$, the metric coefficient functions may be expanded in a power series of the following form
\bqa
&&A = r^2 \sum\limits_{n=0}^{\infty} \f{a_n}{r^{n}},~~B = \sum\limits_{n=0}^{\infty} \f{b_n}{r^{n}},~~\Sigma = r \sum\limits_{n=0}^{\infty} \f{c_n}{r^{n}},\label{equ:seriesexp}
\eqa
where $a_0 = 1$, $b_0 = -\f{2}{3}\log\tau$ and $c_0 = \tau^{\f{1}{3}}$, which are determined by the boundary conditions
\bqa
&&\left.A \right|_{r \rightarrow \infty} = r^2, ~~\left.B \right|_{r \rightarrow \infty} =-\f{2}{3} \log\tau,~~\left.A \right|_{r \rightarrow \infty} = \tau^{\f{1}{3}} r .
\eqa
Specifically,
to the order we will work, we use
\bqa
&A = A_s +\tilde{A}\,,\quad
B=B_s +\tilde{B}\,,\quad
\Sigma=\Sigma_s+\tilde{\Sigma}\,,&\nonumber\\
&\theta = \theta_s  + \tilde{\theta}\,,\qquad
\phi = \phi_s  + \tilde{\phi}\,,&\label{equ:residualmetric}
\eqa
where the index $s$ indicates power series expansions. 
There are only two series coefficients, $a_1$ and $a_4$, that can not be solved from (\ref{Eeqns}). 
Via holographic renormalization~\cite{Balasubramanian:1999re, de Haro:2000xn}, the coefficient $a_4$ is related to the boundary stress tensor by
\bqa
T_{\mu\nu}=- \f{3 \kappa}{4}\text{diag}\left\{ a_4, a_4 + \f{\tau}{ 2 } \partial_\tau a_4, a_4 + \f{\tau}{ 2 } \partial_\tau a_4, - \tau^2 \left( a_4 + \tau \partial_\tau a_4\right) \right\},
\eqa
so that we can read off the energy density, longitudinal and transverse pressure of the medium
as 
$$
\epsilon\equiv - \frac{3\,\kappa\, a_4}{4}\,,\quad
p_L \equiv\frac{3 \kappa}{4}\left( a_4 + \tau \partial_\tau a_4\right)\,,\quad 
p_T\equiv\frac{3 \kappa }{4}\left( -a_4-\f{\tau}{2} \partial_\tau a_4\right)\,.
$$

In contrast,  $a_1$ corresponds to the gauge redundancy in (\ref{equ:diffeomorphism}) and, therefore, does not appear in any physical quantity. 
In appendix \ref{app:A}, all the series metric functions $A_s$, $B_s$, $\Sigma_s$, $\theta_s$ and $\phi_s$ 
needed in our code are given by taking $a_1$ and $a_4$ as arbitrary functions of $\tau$.  
The series expansions in (\ref{equ:seriesexp}) with different gauge choices of $a_1$ are also related to each other according to the transformation in (\ref{equ:diffeomorphism}) and (\ref{equ:metrichat}) with $a_1=2 f$.

\subsection{Numerical method}
\label{sec:alg}

Einstein's equations will be solve by the pseudo-spectral method described in Ref.~\cite{Boyd}: spectral differentiation in $r$ and finite differences in $\tau$.  The algorithm in the simplest gauge choice $f = 0$ is described in details in Ref. \cite{cy2}. In this case, we would 
need to impose lower cutoff $L_{\rm min}$ for the
integration domain which needs to fulfill the requirement $L_{\rm min}<r_h(\tau)$
for all $\tau$. We found this approach to work well for
late time (near-equilibrium) situations, where it is furthermore
computationally cheap. However, at early times (far from equilibrium),
$r_h(\tau)$ depends strongly on $\tau$, and hence it is inconvenient
to set up a computational domain with fixed $L_{\rm min}$. In these circumstances, one can use an alternative method. Since the inside of the horizon is causally disconnected from outside observers on the boundary, the computational domain can be chosen to be $r\ge r_h(\tau)$. One can use the diffeomorphism
(\ref{equ:diffeomorphism}) to fix $r_h$ to a given value, say, unity\footnote{
Careful readers will notice that the mass dimension of $r_h$ would prohibit us to set it to unity.
However, one can fix this by introducing 
an overall dimensionful scale in the problem that will turn out
to cancel everywhere in physical observables. The definitions given 
below should be understood in this sense.}, 
in 
which case the only sensible choice for the cutoff becomes $L_{\rm min}=1$.
Besides the lower cutoff it is also necessary to truncate the computational domain at large radii at $r=L_{\rm max}$ for numerical reasons discussed in the next subsection.

As for any pseudo-spectral method, we have to choose the location
of grid points (corresponding to a choice of basis functions), called
collocation points~\cite{Boyd}. For $N+1$ points we choose
 \beq
 r_j = a e^{b \cos\left( \f{j\pi}{N} \right) }+ c,
 \eeq
 where $a$, $b$ and $c$ are fixed by
  $$ r_0 = L_{max},~~r_N = L_{min},~~\text{and}~~r_{N/2} = H, $$
that is,
 \bqa
 a&=&\f{H^2 - H L_{max} - H L_{min} + L_{max} L_{min}}{2H - L_{max} - L_{min}},\\
 b&=&\log \left( \f{L_{max} - H}{H - L_{min}} \right),\\
 c&=&-\f{H^2 - L_{max} L_{min}}{L_{max} + L_{min} - 2 H}.
 \eqa
 Here, $H$ will be chosen to ensure that our algorithm is numerically stable at a relatively large time step $d\tau$. In the following, we will denote 
 a function $f$ evaluated at any collocation point $r_j$ by $f_j$. Then
the derivative of the function $f$ at $r_i$ is given 
in terms of the differentiation matrix $D_{N+1}$,
$$
f'_i \equiv f^\prime(r_i)=\sum\limits_{j=0}^{N}D_{N+1, ij} f_j\,,\qquad D_{N+1,ij}= \f{e^{-b \cos\left( \f{i\pi}{N} \right)}}{a b} d_{N+1,ij}
$$
 where~\cite{Boyd}
 \beq
         \left( d_{N+1} \right)_{ij} = \left\{
        \begin{array}{c l}
            \f{2 N^2+1}{6}& ~~~~i=j=0\\
            \f{- \cos\left( \f{j\pi}{N} \right)}{2 \left(  1 -\cos^2\left( \f{j\pi}{N} \right) \right)}& ~~~~0<i = j < N\\
           \f{c_i}{c_j} \f{(-1)^{i+j}}{\cos\left( \f{i\pi}{N} \right)  - \cos\left( \f{j\pi}{N} \right)}& ~~~~i\neq j\\
             -\f{2 N^2+1}{6}& ~~~~i=j=N\\           
        \end{array}
        \right.
\eeq
with $c_0=c_n=2$ and $c_j=1$ otherwise. For the sake of numerical accuracy, if 
\beq
\left. f(r,\tau)\right|_{r\rightarrow \infty} = f_0(\tau) r^n,~~n>0~~\text{and~~$f_0$ is a function only of $\tau$},
\eeq
we, instead, calculate the derivative of $f$ by
\beq
f'_i =  \f{n f_i}{r_i} + r_i^n \sum\limits_{i=0}^N D_{N+1,i j} \f{f_j}{r_j^n}.\label{equ:drrefine}
\eeq
Note that since $D_{N+1}$ acting
on a constant vector is vanishing, it contains a zero eigenvalue
and hence is not invertible. For this reason, we consider
an alternative version where we drop the collocation point 
$j=0$ (corresponding to $r=L_{\rm max}$, the point closest to the boundary)
and define an $N \times N$ matrix
\beq
\tilde D_{N,ij}\equiv D_{N+1,ij}\,,\quad i,j=1,\ldots N\,.
\eeq
Using $\tilde D_{N}$ instead of $D_{N+1}$, we have to supply 
boundary conditions at $r=L_{\rm max}$ to conserve the total number of 
equations. $\tilde D_{N}$ is invertible and one can numerically solve $\tilde D_{N}^{-1}$, the inverse of $\tilde D_{N}$, once for all to save computation time.

Assuming that $A$, $B$ and $S$ are known at $\tau$, one can 
first calculate $\theta$ and $\phi$ at $\tau$ by solving (\ref{equ:theta}) and (\ref{equ:phi}) 
with the boundary conditions in (\ref{equ:thetas}) and (\ref{equ:phis}). It is of numerical advantage to deal with the ``residual'' metric functions $\tilde{\theta}$ and $\tilde{\phi}$ defined in (\ref{equ:residualmetric}) instead. Using the differentiation matrices, the solutions are 
\bqa
\label{thetaphisol}
\tilde{\theta}_{j} &=& \sum_{i=1}^{N} \tilde{D}^{-1}_{N, ji} \tilde{S}_{i} \,,\nonumber\\
\tilde{\phi}_{j} &=& 3 \sum_{i=1}^{N}\tilde{D}^{-1}_{N, ji} \left( S_{i}^{-\f{1}{2}} \theta_{i} \sum_{k=0}^N D_{N+1,i k} B_k  - \phi'_{s i} \right),
\eqa
where $j = 1, 2, \cdots, N$, $\tilde{\theta}_{0} = 0$ and $\tilde{\phi}_{0}=0$.
 The numerical algorithm to solve Einstein's equations is then as follows:
\begin{enumerate}
\item
Obtain $a_4$, $S$ and $B$ at $\tau+d\tau$ by solving the 
difference equations of (\ref{equ:a4}), (\ref{equ:dotS}) 
and (\ref{equ:dotB}). To be more specific, in our code the equations are 
solved using a third-order Adams-Bashforth method, that is,
 \beq
 h(\tau+d\tau) = h(\tau)+\f{d\tau}{12}\left[ 23 v(\tau) -16 v(\tau-d\tau) + 5 v(\tau - 2d\tau)\right],
 \eeq
 for a general ordinary differential equation of the form $ \f{dh}{d\tau} = v$.
\item
Then, calculate $b_4$ and $a_1/r_h$ at $\tau+d\tau$. To do this, we use the  first-order implicit Euler scheme to discretize
\beq
\f{d}{d\tau}h(\tau+d\tau) = \f{h(\tau+d\tau)-h(\tau)}{d\tau},~~\f{d^2}{d\tau^2} h(\tau+d\tau)= \f{h(\tau+d\tau)-2h(\tau)+h(\tau-d\tau)}{d\tau^2},\label{equ:a1dis}
\eeq
where $h=a_1,b_4$.
In this paper, we use the following two algorithms corresponding to two different ways to fix the gauge function $f=a_1/2$.
\begin{enumerate}
\item Alg. I: $f = 0$\\
In this case, one needs only to solve $b_4$ from the discretized version of the 
equation $\left.B_{s0}\right|_{a_1=0}=B_0$, and 
$r_h(f=0)$ can be calculated by $\theta(r_h,\tau+d\tau)=0$
after one has obtained $\theta$ at $\tau+d\tau$ in Step 3.

\item Alg. II: $r_h(f) = 1$\\
In this case, one needs to solve two coupled differential equations given by
$B_{s0}=B_0\label{equ:a1b4a}$ and 
$\theta_N(\tau + d\tau)\equiv \theta(1,\tau+d\tau)=0$.
Using the discretization in (\ref{equ:a1dis}), one can express
$b_4(\tau+d\tau)$ as a function of $a_1(\tau+d\tau)$ 
and solve $\theta_N(\tau + d\tau)=0$ for $a_1(\tau+d\tau)$ 
using  (\ref{thetaphisol}) and (\ref{equ:thetas}).
\end{enumerate}

\item
Next, calculate $\theta$ and $\phi$ at $\tau + d\tau$ by (\ref{thetaphisol}) with the boundary conditions given by $\theta_0=\theta_{s0}$ and $\phi_0=\phi_{s0}$, or equivalently,  $\tilde{\theta}_0=0$ and $\tilde{\phi}_0=0$.
\item
Finally, one can calculate $A = \tilde{A} + A_s$ at $\tau+d\tau$ by integrating (\ref{equ:App}). The boundary conditions are given by $A_0=A_{s0}$ and $A'_0=A'_{s0}$, or equivalently, $\tilde{A}_0 = 0$ and $\tilde{A}'_0=0$.  $D_{N+1}^2$ is not invertible either because it has two eigenvectors with eigenvalue 0.  As a result, the discretized equation of (\ref{equ:App}) gives us only $N-1$ linearly independent equations, which can be chosen as
\bqa
 \sum_{j=1}^{N} D^2_{N+1, i j} \tilde{A}_j &=&   \left( 8  \theta_i S_i{'} S^{-2}_i + 3  \phi_i S^{-\f{1}{2}}_i  \sum_{j=0}^{N} D_{N+1, i j}  B_j\right) \nonumber\\
&-& 4- A{''}_{si} ,\label{equ:Appd}
\eqa
where $i = 2, ..., N$ and $S{'}$ is calculated from (\ref{equ:drrefine}) with $n = 3$. One needs one more equation to solve all $A_i$ with $i = 1, 2, \cdots, N$, which is given by 
\beq
\tilde{A}'_0= \sum_{i=1}^{N} D_{N+1, 0 i} \tilde{A}_i = 0.\label{equ:Apd}
\eeq
$\tilde{A}$ at $\tau+d\tau$ can be easily obtained by solving the $N$ linear equations (\ref{equ:Appd}),(\ref{equ:Apd}).
\item
Repeating steps 1-4 one can get the geometry in the bulk at all times. 
\end{enumerate}
To summarize, the numerical algorithm discretizes Einstein's equations using the calculational
parameters $L_{\rm max},N,d\tau$ and fixing either $f=0,L_{\rm min}$ 
or $r_h(f)=1$, $L_{\rm min}=1$. The continuum
Einstein's equations are recovered in the limit $L_{\rm max}\rightarrow \infty,N \rightarrow \infty,
d\tau\rightarrow 0$.

From the description above one can expect that the algorithm in the gauge choice $f = 0$ (Alg.~I) 
should be computationally cheaper than that with $r_h(f) = 1$ (Alg.~II). 
At late times, the location of the apparent horizon $r_h(f=0)$ approaches $r=0$. Since $L_{min}<r_h$, this implies choosing $L_{min}\sim 0$. However, this can not always be done. We shall see in the next section that for some values of $\bar c$ there are coordinate singularities at $r=r_V\sim r_h$ in the initial metric set up by (\ref{sigmaansatz}). In this case, one has to choose  $L_{min}\geq r_V$ and, as a result, to stop the code when $r_h$ falls below $L_{min}$. In the following, we will use both algorithms: Alg. I for the cases $r_V \ll r_h(\tau_0)$ and Alg. II for the cases $r_V \sim r_h(\tau_0)$.

\subsection{Code tests: late time dynamics}

Let us present the performance of the above algorithms in the case where
analytic results are available: the late time (hydrodynamic) behavior.
In this case, the initial conditions for the code are given 
by the following approximate solutions in the gauge $a_1 = 0$
\beq
A = r^2 + \f{a_4}{r^2},~~B = - \f{2}{3} \log\left( \tau + \f{1}{r} \right),~~\text{and}~~S = \tau r^3 + r^2,\label{equ:latetimemetric}
\eeq 
where $a_4$ in the leading-order in $\f{1}{\tau}$ is given by 
$a_4 = - \f{w_0^4}{\tau^{4/3}}$
with $w_0$ a constant. It should be emphasized that this initial condition 
is only an approximate solution to Einstein's equations, whereas
in the later part of this work we will work with exact solutions
as initial conditions. Here, we will investigate
the time evolution of the error thus made.

\begin{figure}
\begin{center}
\includegraphics[width=0.45\linewidth]{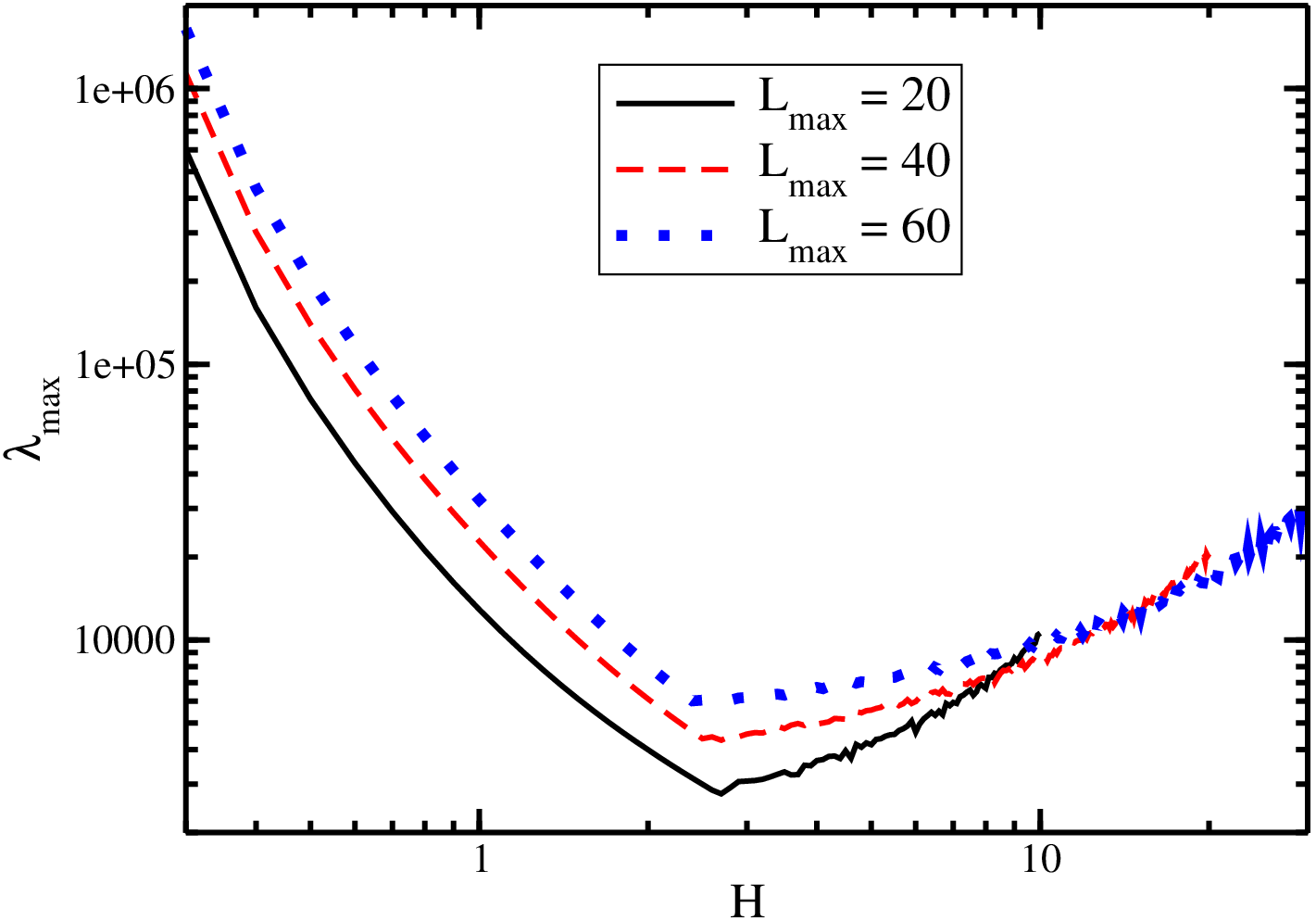}
\hfill
\includegraphics[width=0.45\linewidth]{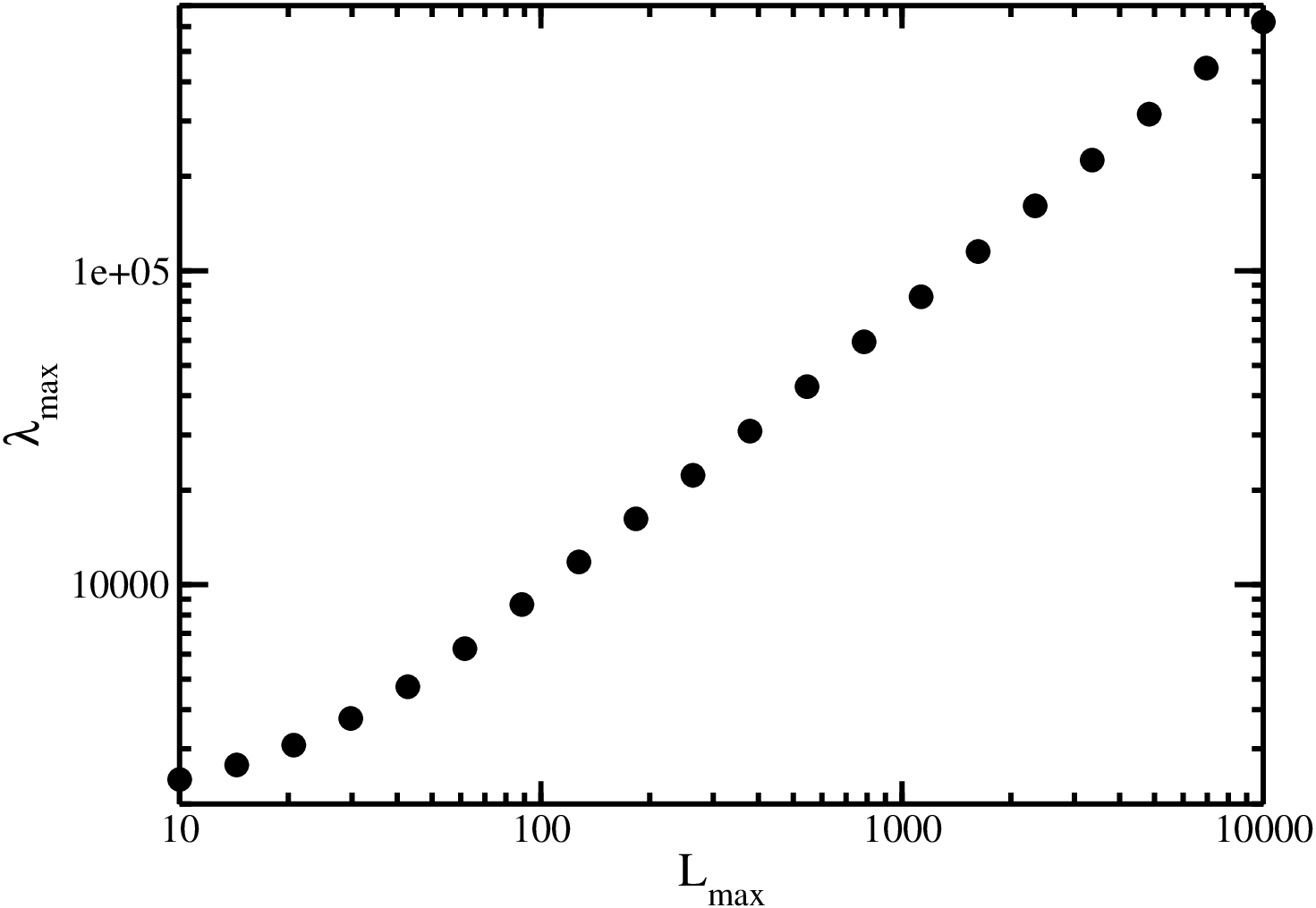}
\end{center}
\caption{The maximum eigenvalue $\lambda_{\rm max} $  
in Alg. I for $N = 128$, $L_{\rm min} = 0.2$ and $w_0^{3/2}\tau = 4$. Left:
the dependence of $\lambda_{\rm max}$ on $H$. For $L_{max} = 20, 40$ and $60$,
the minimum $\lambda_{\rm max}$ is found at $H = 2.7, 2.7$ and $2.6$, respectively.
Right: dependence on $L_{max}$ with $H = 3.0$.
}\label{fig:stability}
\end{figure}

Performing a full-blown numerical stability 
analysis of our algorithm would be interesting, but rather complicated,
so we leave it for future work. However, an approximate stability criterion can be found 
by considering the differential 
operator on the left-hand side of (\ref{equ:dotS}) and (\ref{equ:dotB}), 
that is, $\partial_\tau - \f{1}{2} A \dr$. Discretizing this operator we find
\beq
\f{1}{d\tau} \left( \delta_{ij} + \f{1}{2} d\tau O_{ij} \right),~~O_{ij} \equiv A_i D_{N+1,ij}.
\eeq 
One can argue that the for the algorithm to be stable, the time increment $\delta\tau$
has to be small enough that $d\tau O_{ij}<\delta_{ij}$. Estimating the size of $O_{ij}$
by its maximum eigenvalue $\lambda_{\rm max}$, we find
$
d\tau \lesssim \f{1}{ \lambda_{max}}\,.
$

As shown in Fig. \ref{fig:stability}, the maximum eigenvalue $\lambda_{\rm max}$ depends
on the choice of the parameter $H$ as well as $L_{\rm max}$. 
From this figure, one can see that the choice $H=3$ effectively minimizes $\lambda_{\rm max}$
and hence should allow algorithmic stability for larger time increments $d\tau$.
We adopt this choice in following. With $\lambda_{max}\sim 5\times 10^{3}$
we therefore expect $d\tau \lesssim 10^{-4}$ to be necessary for algorithmic stability.

\begin{figure}
\begin{center}
\includegraphics[width=0.41\linewidth]{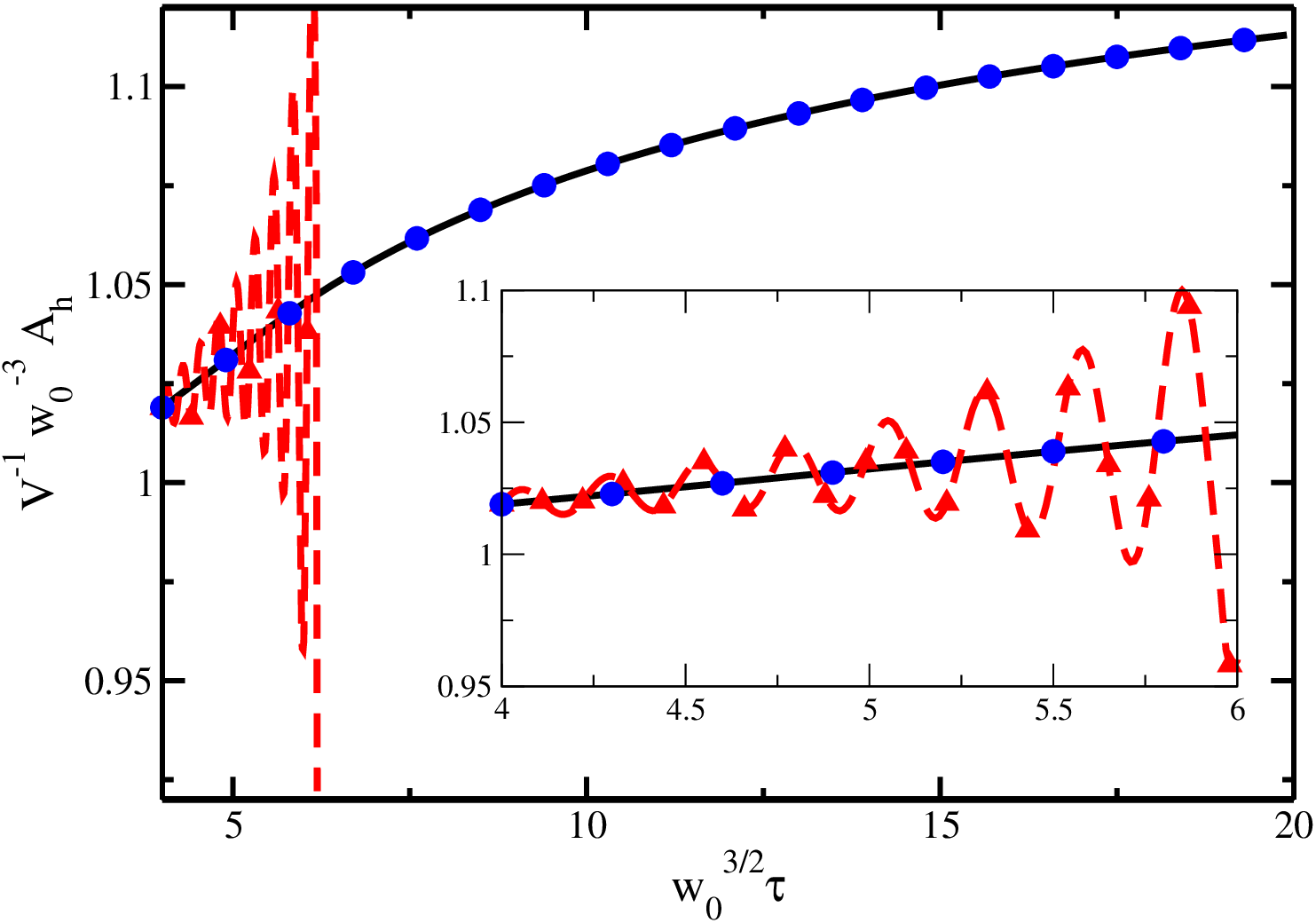}\hfill
\includegraphics[width=0.41\linewidth]{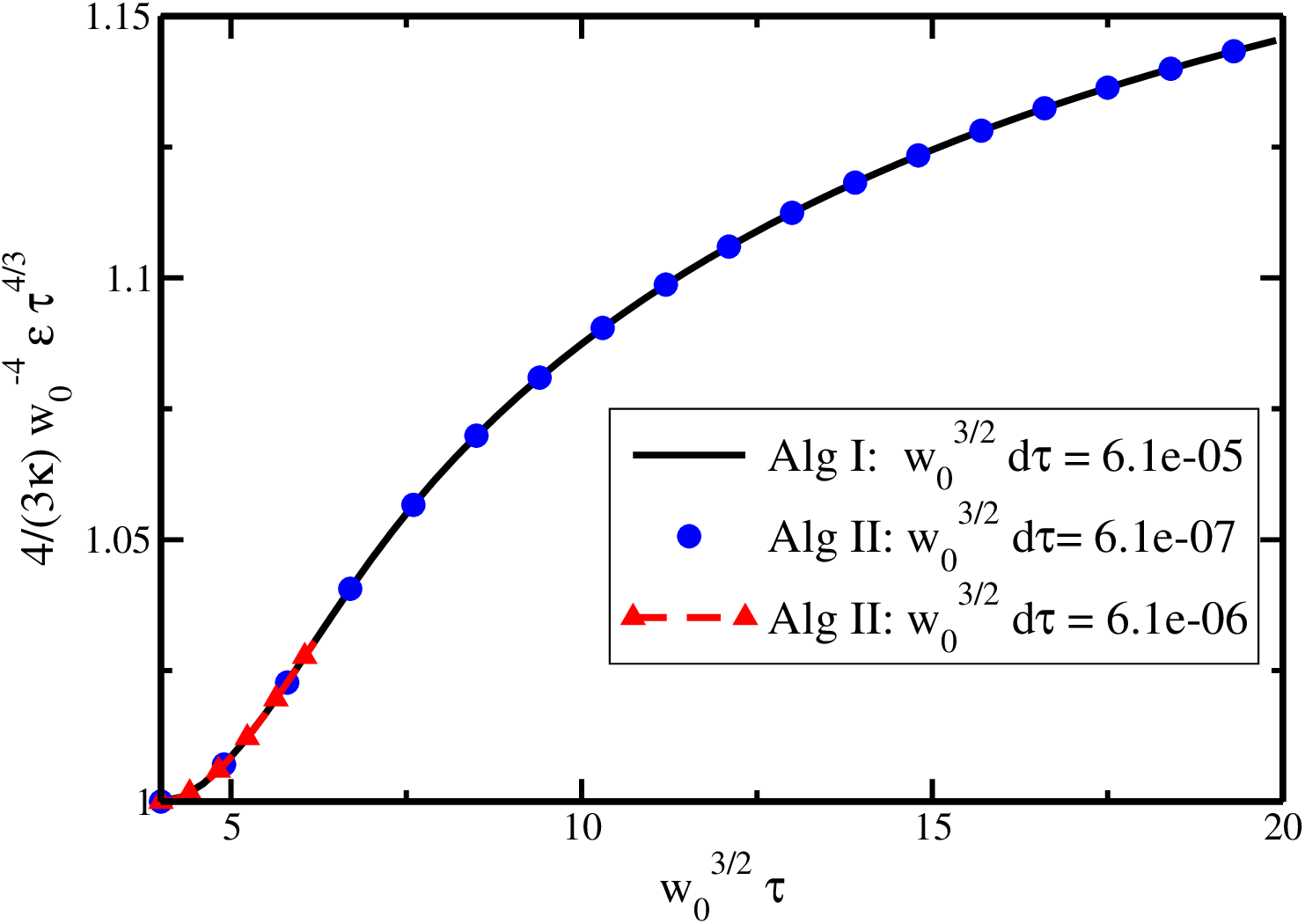}
\end{center}
\caption{Numerical stability of the two algorithms: gauge choice $f=0$ (denoted Alg.~I) and
gauge choice $r_h(f)=1$ (Alg.~II). Shown are results for the horizon area $A_h$ (left, inset
zooms to early time behaviour), 
and the energy density $\epsilon$ (right) 
for $L_{max} = 20$, $N = 128$ and $H = 3.0$.
For Alg.~I, results for one choice of time increment $d\tau$ are shown 
(further decreasing $d \tau$ leaves the result unchanged). 
In order to achieve stable results for Alg.~II,
we have to decrease $d \tau$ considerably. When this is done, the results
from Alg.~II match those from Alg.~I.}\label{fig:alg}
\end{figure}

In Fig. \ref{fig:alg}, we show results by Alg. I with $d\tau =6.1\times 10^{-5}$ and Alg. II with $d\tau =6.1\times 10^{-6}$ and $6.1\times 10^{-7}$. 
Using Alg. II with $d\tau =6.1\times 10^{-6}$, we get numerically unstable results
for the horizon position and area.
However, by choosing a smaller time step,
numerical stable results can also be obtained by Alg. II, 
which agree very well with those by Alg. I. This provides evidence for the equivalence of the two algorithms. 
As a rule of thumb, we find that $d\tau = \f{1}{N^2}$ for Alg. I and 
$d\tau = \f{10^{-2}}{N^2}$ for Alg. II generally ensures numerical stability.

For numerical reasons, it is difficult to do simulations for $L_{\rm max}\gtrsim 100$.
This can be understood from Fig. \ref{fig:stability}, where it is shown that 
$\lambda_{max}$ increases exponentially with $L_{max}$, forcing a similar decrease
in the time increment $d\tau$.
Also, for large $L_{max}$, one confronts the subtraction of two nearly equal numbers in step 3
of the algorithm outlined in Sec. \ref{sec:alg}. Fortunatly, we find in practice
that for $L_{\rm max} w_0^{3/2} \tau_0 \gtrsim 10$ our numerical results stabilize and
do not change appreciably when further increasing $L_{\rm max}$. Thus, we 
are confident that the results reported in the following 
are close to the continuum limit $L_{\rm max}\rightarrow\infty$.
Conversely, note that for $w_0^{3/2}\tau_0\lesssim 0.1$ we would need
$L_{\rm max}\gtrsim 100$ and therefore cannot report results for very early initial times.

We have also studied the dependence of our results on the number of 
collocation points $N$. We find that results for $N=64,128,256$ with $d\tau=1/N^2$
are essentially indistinguishable, while $N=32$ is numerically unstable for $d\tau=1/N^2$,
and differs on the percent level for $d\tau=0.1/N^2$. Thus, we are confident
that the choice $N=128$ is sufficiently close to the continuum $N\rightarrow \infty$ result
and shall adopt this choice in the following.

\begin{figure}
\begin{center}
\includegraphics[width=0.5\linewidth]{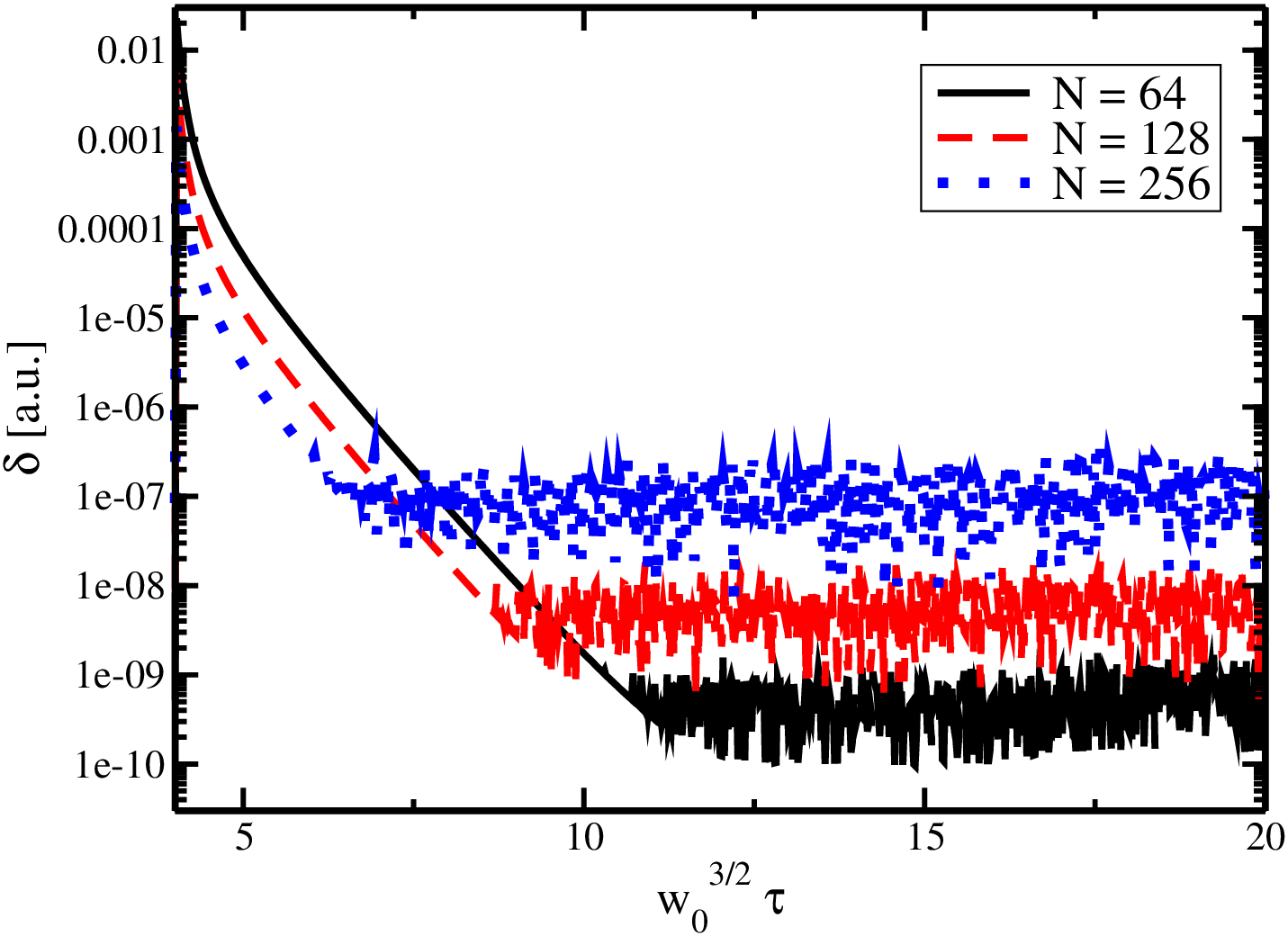}
\end{center}
\caption{Numerical error $\delta$ in the constraint equation (\ref{equ:eee})
as a function of $\tau$ for $N = 64, 128$ and $256$. To test the code,
we start from initial conditions that do not fulfill Einstein's equations, 
so $\delta$ is initially large, but we find that 
in all cases $\delta$ decreases rapidly 
as a function of time.
}\label{fig:constraint}
\end{figure}

Since the initial geometry specified in Eq.~(\ref{equ:latetimemetric}) is only
an approximate solution to Einstein's equations, it is important to check whether
time evolution will decrease or increase the error. 
To answer this question quantitatively, we investigate the constraint equation (\ref{equ:eee})
by defining at each $\tau$
\beq
\delta \equiv \underset{\{r\}}{\text{max}}\left|  \dr^2\Sigma + {\textstyle \frac{1}{2}} B'^2 \, \Sigma\, \right|\,,
\eeq
where for an exact solution to Einstein's equations $\delta=0$.
As shown in Fig. \ref{fig:constraint},  
$\delta$ initially is sizeable but decreases as a function of time until
eventually stabilizing several orders of magnitude below its initial value. 
This implies that our algorithm approaches the exact solution to Einstein's equations
as time advances, rather than further deviating from it.
A physical interpretation is as follows: approximate solutions 
satisfy Einstein's equations at large $r$,
but not close to the horizon position $r_h$. However, in the particular coordinates
we have chosen, the black hole acts as an absorber of the 'offending' modes, pulling
them behind the horizon.
As a result, the approximate solutions can quickly converge into exact solutions.


\begin{figure}
\begin{center}
\includegraphics[width=0.32\linewidth]{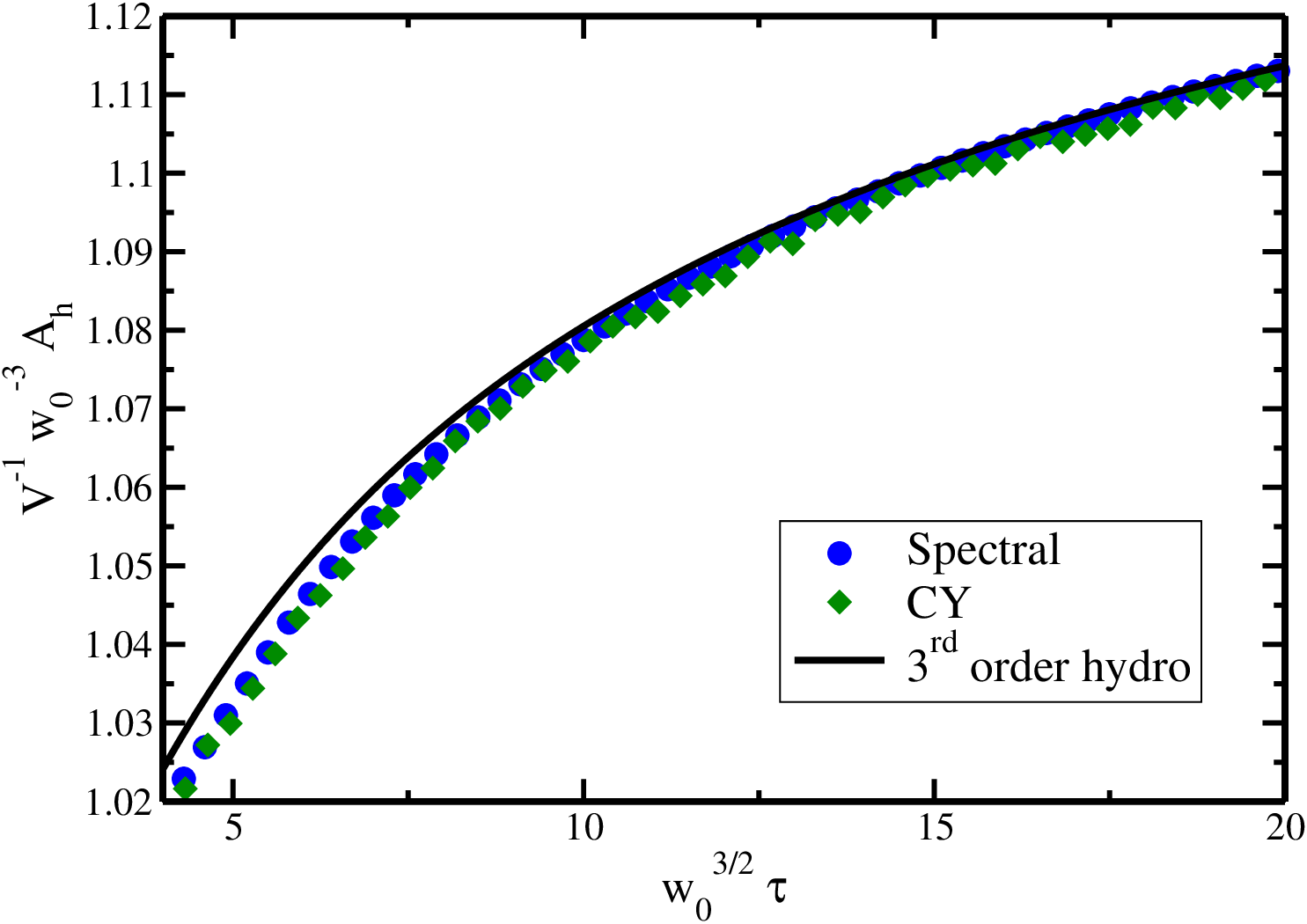}
\includegraphics[width=0.32\linewidth]{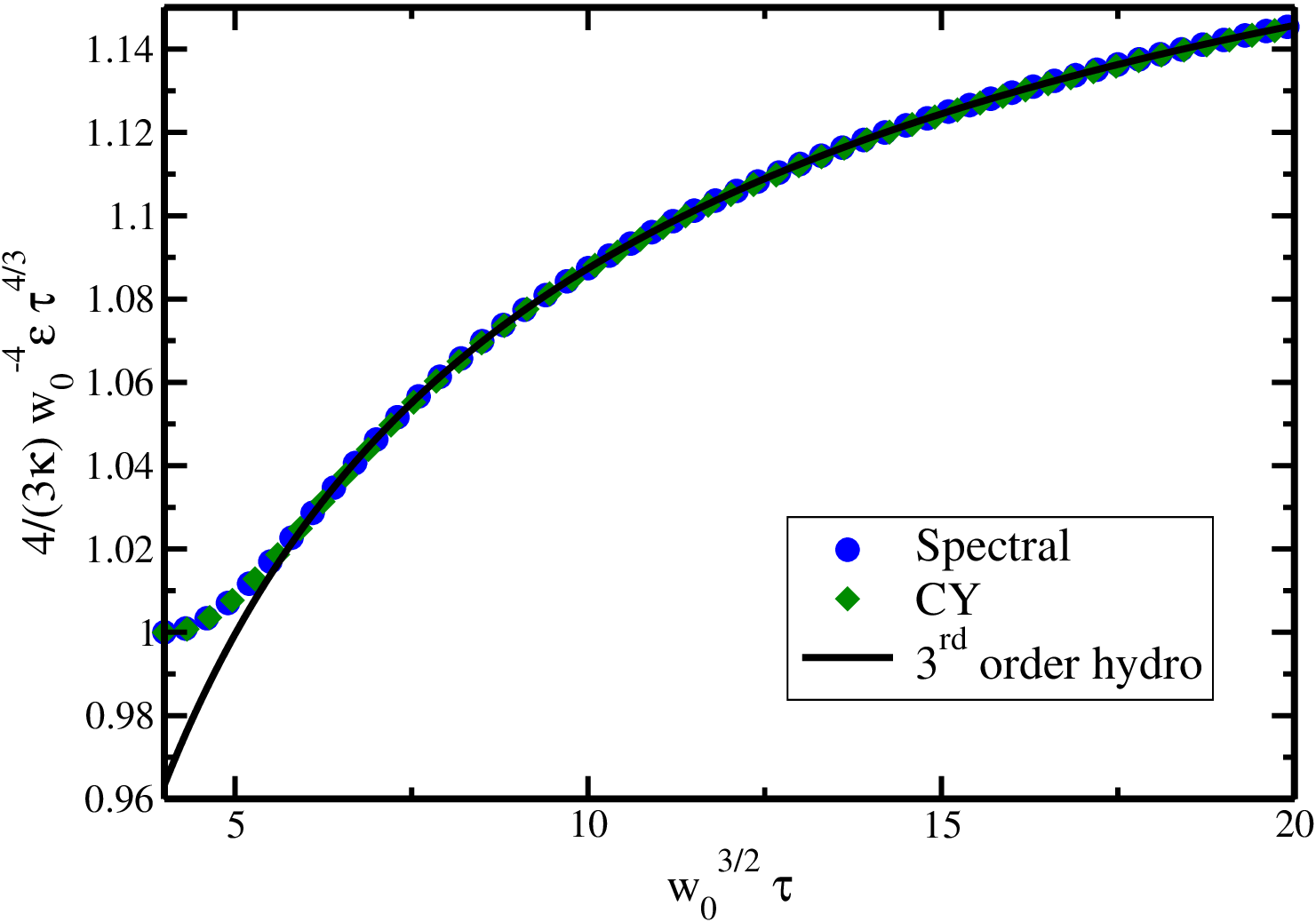}
\includegraphics[width=0.32\linewidth]{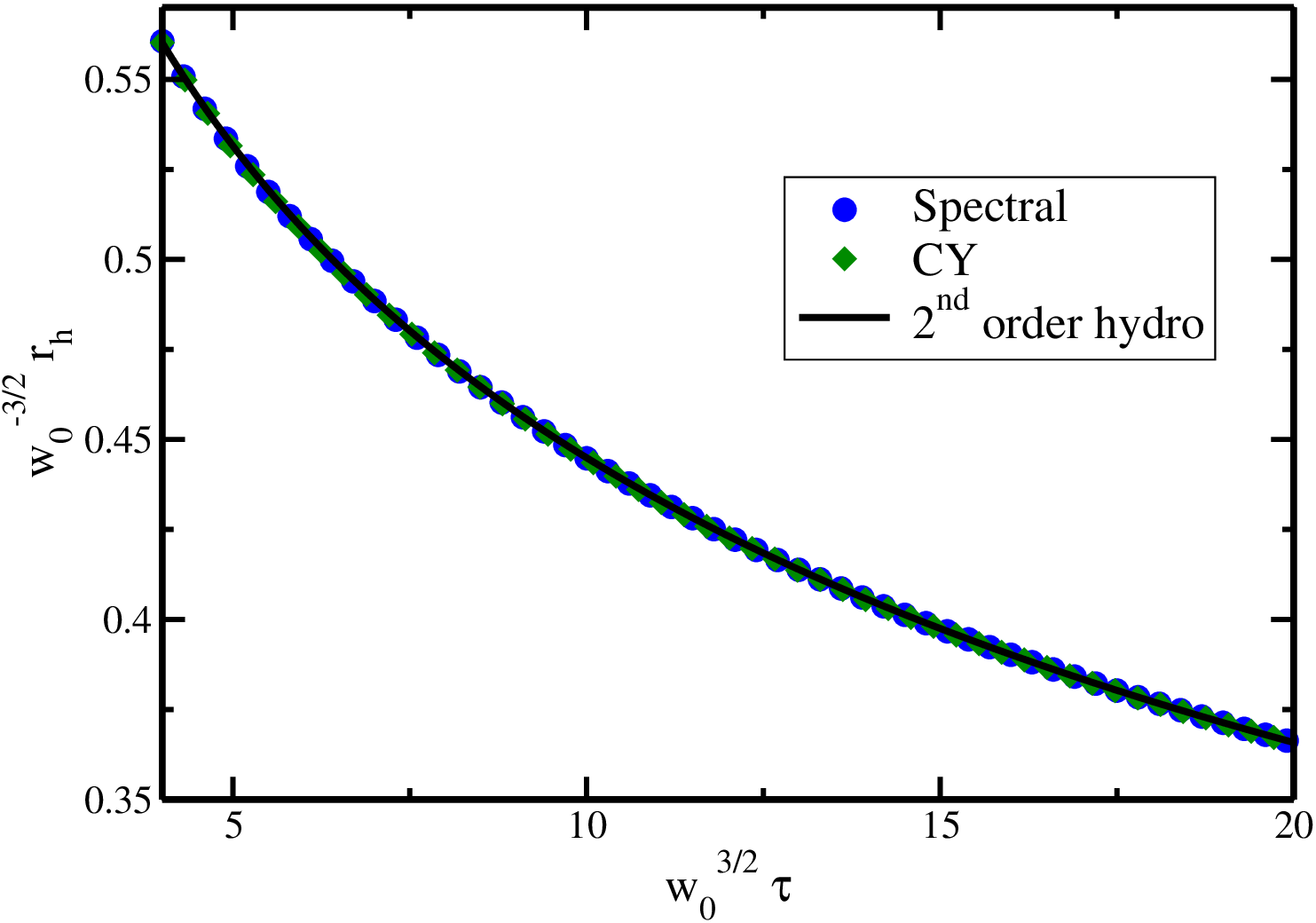}
\end{center}
\caption{Comparison of numerical results from our algorithm (``Spectral''), the algorithm from Ref.~\cite{cy1} (``CY'') and the $2^{nd}$/$3^{rd}$ order hydrodynamics. 
Here, we use $\tau_0 w_0^{3/2} = 4$, $L_{max} = 20$, $N=128$, $d\tau = \f{1}{N^2}$. 
Shown are $r_h(f=0)$ (right), energy density $\epsilon$ (center) and apparent horizon
area $A_h$ (left). Performing a least-square fit of the numerical result for $\epsilon$ with 
hydrodynamics we get $w = 1.0573\, w_0$. The comparison between numerics and hydrodynamics 
for $A_h$ and $r_h$ (left,right) uses this value of $w$.
}\label{fig:cmpwithcy}
\end{figure}

The late time hydrodynamic results for $\epsilon, A_h$
and $r_h$ are given by~\cite{Booth:2009ct}
\bqa
\frac{4 }{3 \kappa}\epsilon^{\rm hydro}&=&\frac{w^{4}}{\tau^{4/3}}-\frac{2 w^3}{3 \tau^2}+\frac{ 1+2 \log(2)}{18 \tau^{8/3}}w^2\nn\\
&+&\frac{ -3+2 \pi^2+24 \log(2)-24 \log^2(2)}{486 \tau^{10/3}}w+{\cal O}(\tau^{-4})\,,\label{equ:a4hydro}\\
V^{-1}A_h^{\rm hydro}&=&w^3-\frac{ w^2}{2 \tau^{2/3}} + \f{2+\pi+6 \log(2)}{24 \tau^{4/3}} w \nn\\
&+&\frac{\pi^2-60 \left(-1+\log(2)+12 \log(2)^2\right)+18 \pi (1+6 \log(2))}{2592 \tau^2}+{\cal O}(\tau^{-\f{8}{3}}),\\
r_h^{\rm hydro}&=& \f{w}{ \tau^{1/3}} - \f{1}{2 \tau} + \f{8+3 \pi - 4 \log(2)}{72 w \tau^{5/3}}\label{equ:Ahhydro}\nn\\
&+& \frac{1}{w^2 \, \tau ^{7/3}} \left(\frac{\mathcal{C}}{18 }+\frac{\pi w
  \delta_3}{3}-\frac{25\pi}{432}+\frac{1}{81}-\frac{\pi^2}{7776}+\frac{7 
  \log^2(2)}{162} + \right. \nonumber \\ 
&-& \left. \frac{\pi \log (w)}{18}-\frac{2
   \log (w)}{27}-\frac{25 \log(2)}{162}\right)+{\cal O}(\tau^{-3})\label{equ:rhhydro},
\eqa
where $r_h$ corresponds to the gauge choice $f = 0$
\footnote{There is one integral constant $\delta_3$ in $r_h$ which we can not fix because the $3^{rd}$-order hydrodynamic formula of $A$ is still missing in the literatures.} and $\mathcal{C}$ is Catalan's constant.
To test the accuracy of our code, we show
the comparison between our algorithms, the results from Ref.~\cite{cy1}
and hydrodynamics in Fig.~\ref{fig:cmpwithcy}. 
Note that
there is very good agreement between the algorithm Ref.~\cite{cy1} and the code used
in this work. We also find that  
the energy density matches the $3^{rd}$ order hydrodynamic result at $\tau \gtrsim 6$. 
We extract the parameter $w$ governing the hydrodynamic behaviour (\ref{equ:a4hydro})
by performing a least-square fit to our numerical result for $\epsilon$. Using this value of $w$,
$A_h$ matches hydrodynamics at a relatively late time 
while the location of the apparent horizon $r_h$ does so at comparatively earlier times. We recall that the initial conditions we had chosen
did not fulfill Einstein's equations (see Fig.\ref{fig:constraint}),
yet at late times, we recover hydrodynamics with the correct expansion
coefficients (\ref{equ:a4hydro},\ref{equ:Ahhydro},\ref{equ:rhhydro}).
For exact initial conditions, we expect even better performance.

To summarize, we have extensively tested our numerical algorithm
 by studying the dependence of the results on the numerical parameters
$L_{\rm max},N,d\tau$, suggesting that we can indeed extract results corresponding
to solutions of the continuum Einstein's equations unless starting at very
early times $\tau$. At late times, our numerical results match the 
analytically known hydrodynamic behaviour as well as those from an independent code.
In the following, we will now use this algorithm to numerically calculate 
the solution to Einstein's equations for initial conditions modelling the collision of shock waves.

\section{Results}\label{sec:results}

In this section, we study the toy model described in Sec. \ref{sec:metricansatz} using the algorithms in the previous section. Using holography, this corresponds 
to a boost-invariant medium with energy density 
$\epsilon(\tau_0\ll 1) = \kappa \mu^2 \tau_0^2$. The increase of $\epsilon$
mimics the 'contracting' stage of two nuclei passing through each other in heavy ion collisions. 
At late times, the system is expected to be described by hydrodynamics. 
Since the system expands along the longitudinal direction, 
one can expect that the medium has to stop contracting, and $\epsilon$ should
eventually decrease in order to match onto the hydrodynamic behaviour.
Within our toy model, we are able to follow and study all stages of this evolution 
quantitatively in a strongly coupled ${\cal N}=4$ SYM medium.

\subsection{Initial conditions}
\label{sec:ICs}

Using the ansatz metric function in (\ref{sigmaansatz}), we set up the initial geometry 
at time $\tau=\tau_0$ by the steps given below. \textit{Note that --- unlike the test case
considered in the previous section --- the resulting initial condition is an exact 
solution to Einstein's equations.}

\begin{enumerate}
\item Initialize $a_4$. 
Near the boundary $r\rightarrow \infty$, the power series expansion of our ansatz metric function $\Sigma$ in (\ref{sigmaansatz}) is the same as 
that of $\Sigma$ in (\ref{equ:metricshockapp}) up to ${\cal O}(\f{1}{r^7})$. Therefore, $a_4$ must be given by the same expression as that of the approximate 
solutions in (\ref{equ:metricshockapp}), that is, 
$a_4(\tau_0) = -\f{4\bar \mu^2 \tau_0^2}{3}$.

\item Then, initialize $a_1(\tau_0)$, $a_1(\tau_0-d\tau)$ and $a_1(\tau_0 -  2d\tau)$.
Inserting the ansatz in (\ref{sigmaansatz}) into (\ref{equ:theta}), we can solve $\theta$ analytically in the gauge $f=\frac{a_1}{2} = 0$,
\bqa
\theta&=&\frac{r^3}{3}+\frac{r^4\tau}{4}-\frac{\mu \pi (36+35 \sqrt{\bar c}\bar\mu^{5/3}\tau^2)}{
105 \sqrt{2} \bar c^{3/4}}\nonumber\\
&&-\frac{\bar\mu \left(36 \sqrt{2}+7 \bar c^{1/4} \bar\mu^{1/3} \tau  \left(18+5 \sqrt{2} \bar c^{1/4} \bar\mu^{1/3} \tau \right)\right) }{210 \bar c^{3/4}}
\text{arctan}\left[1-\frac{\sqrt{2} r}{\bar c^{1/4} \bar\mu^{1/3}}\right]\nn\\
&&+\frac{\bar\mu \left(36 \sqrt{2}+7 \bar c^{1/4} \bar\mu^{1/3} \tau  \left(-18+5 \sqrt{2} \bar c^{1/4} \bar\mu^{1/3} \tau \right)\right) }{210 \bar c^{3/4}}
\text{arctan}\left[1+\frac{\sqrt{2} r}{\bar c^{1/4} \bar\mu^{1/3}}\right]\nn\\
&&+\frac{\sqrt{2} \bar\mu \left(-36+35 \sqrt{\bar c} \bar\mu^{2/3} \tau ^2\right)}
{420\bar c^{3/4}} 
\left(\text{log}\left[\frac{
\sqrt{\bar c} \bar\mu^{2/3}+r^2-\sqrt{2}\bar c^{1/4} \bar\mu^{1/3} r}
{\sqrt{\bar c} \bar\mu^{2/3}+r^2+\sqrt{2}\bar c^{1/4} \bar\mu^{1/3} r}\right]
\right).
%
%
\eqa

Under the transformation (\ref{equ:diffeomorphism}) with $f=a_1/2$ we get
$\theta(f)$ and therefore can obtain 
$a_1$ by solving $\theta=0$. In Alg. I, one can skip this step.

\item Next, initialize $b_4(\tau_0)$, $b_4(\tau_0 - d\tau)$ and $b_4(\tau_0 - 2 d\tau)$. As for $a_4$, one can get $b_4$ 
from the power series expansion of $B$ in (\ref{equ:metricshockapp}). In this way, we have 
\beq
b_4 = \f{2+20 \bar \mu^2 \tau^6+\tau a_1 (4+\tau a_1 (3+\tau a_1))}{12 \tau^4}.
\eeq

\item Initialize $B(\tau)$ by integrating 
(\ref{equ:eee}) with boundary conditions given by (\ref{perturbative}).
 
 \item Finally, obtain $\theta(\tau_0)$, $\phi(\tau_0)$ by (\ref{thetaphisol}) and $A(\tau_0)$ by solving the linear equations in (\ref{equ:Appd}) and (\ref{equ:Apd}).

\end{enumerate}
To use the third-order Adams-Bashforth method we also calculate the metric functions at $\tau_0 - d\tau$ and $\tau_0 - 2d\tau$ by repeating all the steps above.

As explained in the previous section,
it is not possible to choose $\bar\mu^{1/3}\tau_0=0$ for numerical reasons. 
However, since the initial 
conditions become less and less reliable for larger $\tau_0$, we 
want to choose $\bar\mu^{1/3}\tau_0$ as small as possible such that
the numerical algorithm can still be applied. The smallest
value we achieved in practice was $\bar\mu^{1/3}\tau_0=0.2$.

\subsection{Transition to hydrodynamic behaviour}

\begin{figure}
\begin{center}
\includegraphics[width=0.45\linewidth]{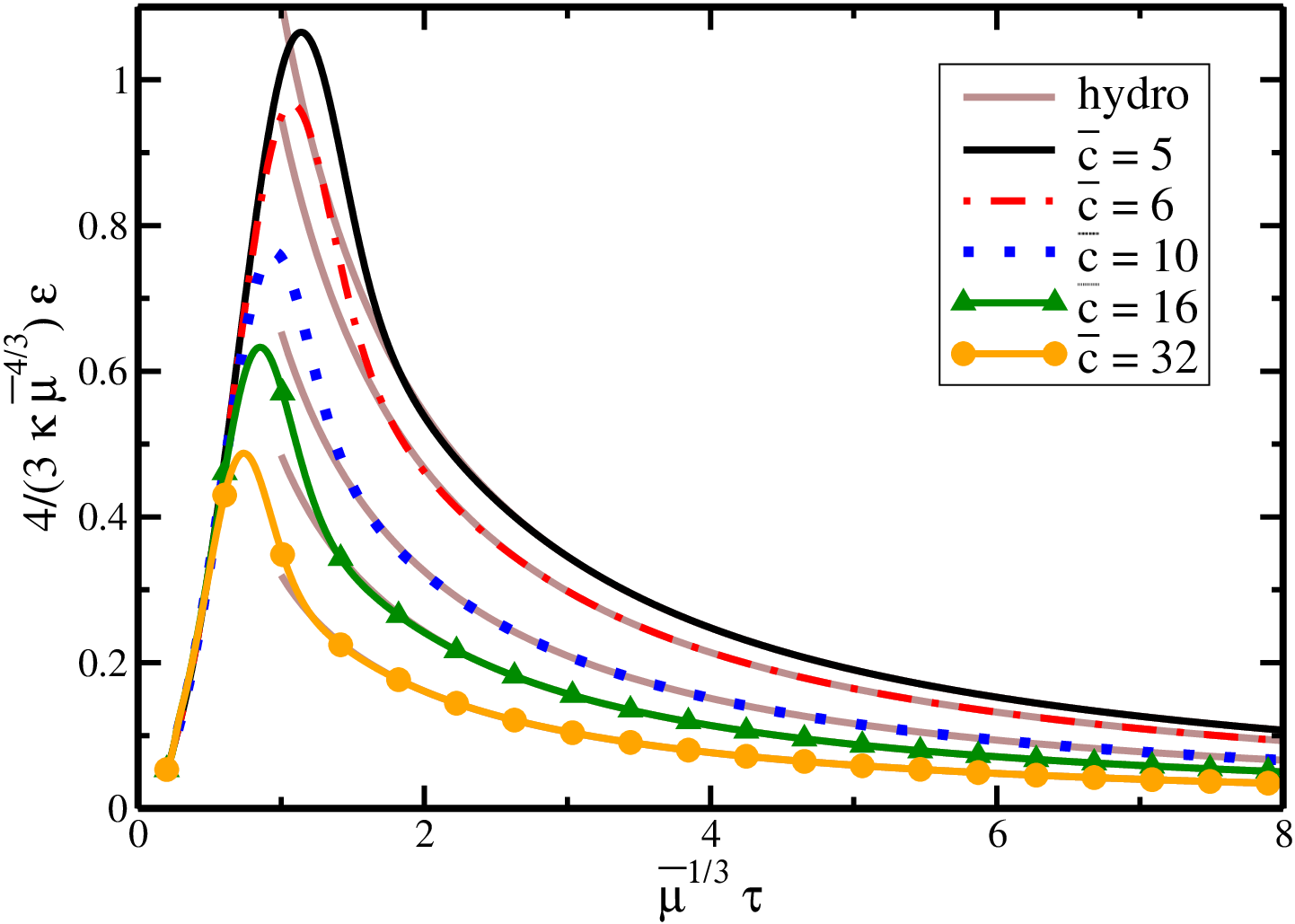}\hfill
\includegraphics[width=0.5\linewidth]{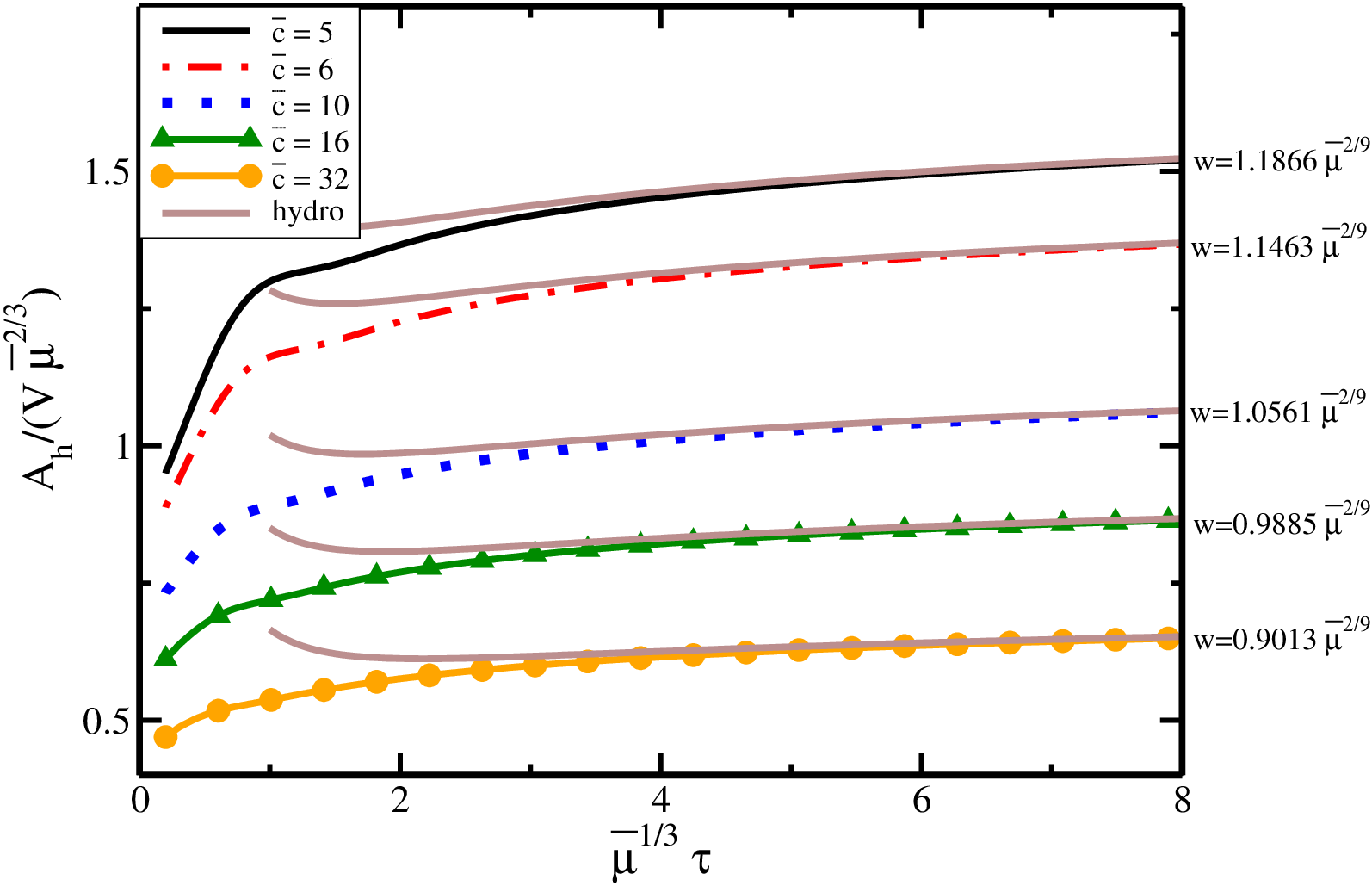}
\end{center}
\caption{The $\tau-$evolution of the energy density and horizon area
for $\tau_0 \bar \mu^{1/3} = 0.2$ and different values 
of $\bar c$. 
Lines labelled 'hydro'
represent $3^{rd}$ order hydrodynamic results. 
The $w$ values indicated are obtained from hydrodynamic fits to $\epsilon$.
}\label{fig:tau20}
\end{figure}

The initial conditions we consider do not exhibit hydrodynamic
behaviour at early times. This can be clearly seen from the 
time dependence of the energy density, Eq.(\ref{ED}), which 
is very different from the hydrodynamic $\tau^{-4/3}$ result.
To study the transition from the early time behaviour to hydrodynamics,
we choose particular values for $\tau_0,\bar{c}$ and 
then evolve the initial conditions in Sec. \ref{sec:ICs} forward
in time using our numerical algorithm.

In Fig. \ref{fig:tau20}, the apparent horizon area and energy density 
are shown for initial conditions with $\bar \mu^{1/3}\tau_0=0.2$
and various values of $\bar c$. As can be seen from this figure, the energy density
first increases, reaches a maximum at around $\bar \mu^{1/3}\tau_0\simeq1$, and then 
starts to decrease. One expects the late time dynamics to be described by 
hydrodynamics, ~Eq.(\ref{equ:a4hydro}). 
We perform a hydrodynamic least-square fit to our results for $\epsilon$ to extract the 
parameter $w$ at times\footnote{We have checked that the extracted
values for $w$ change 
by less than $0.0011\%$
if we perform the fit for $\bar \mu^{1/3}\tau>6$, indicating the insensitivity of the extracted
$w$ values.}
$\bar \mu^{1/3}\tau>7.5$. The hydrodynamic results are shown together with
the full numerical results in Fig.~\ref{fig:tau20}. As can be seen,
the numerical late-time behaviour of both the energy density as well
as the horizon area (using the same $w$ values) are very well described by hydrodynamics
for all chosen values of $\bar c$. One should note that the energy density is initially 
independent of $\bar c$, but its late time behaviour differs for different $\bar c$.
This indicates that after some pre-equilibrium
stage, the system indeed thermalizes, with the overall scale $w$ dependent
on the non-equilibrium initial conditions.

\begin{figure}
\begin{center}
\includegraphics[width=0.43\linewidth]{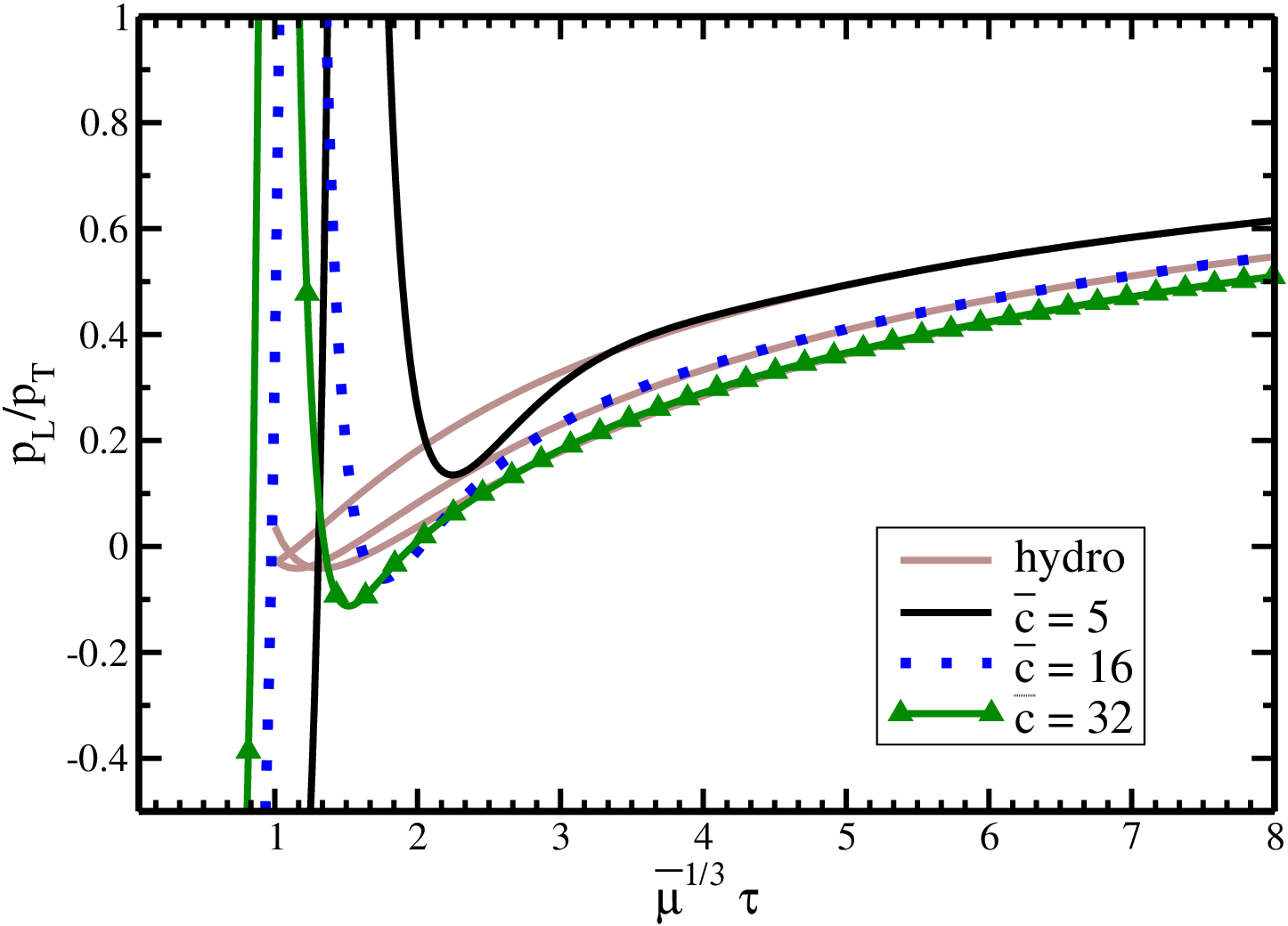}\hfill
\includegraphics[width=0.45\linewidth]{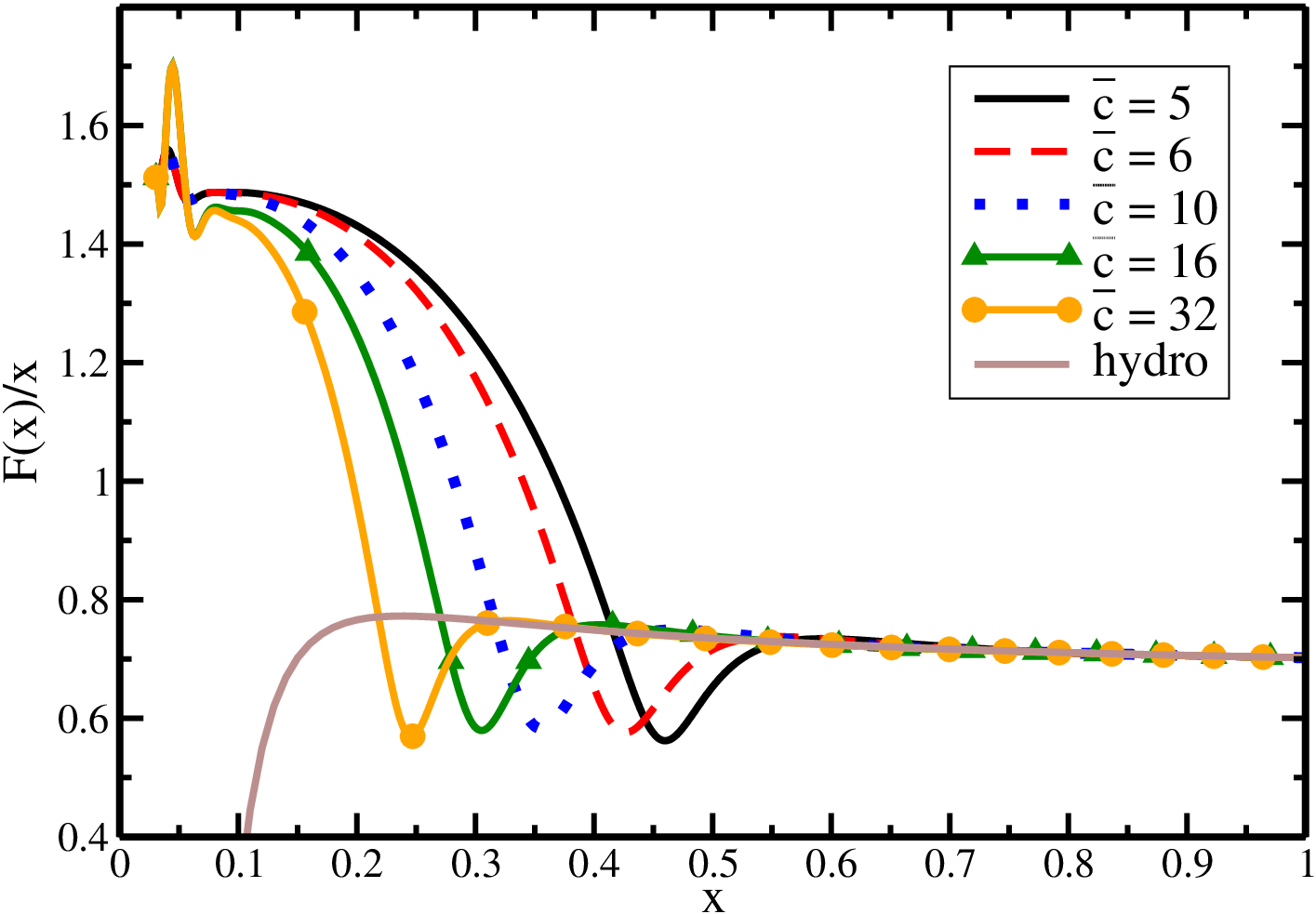}
\end{center}
\caption{
Left: The $\tau-$evolution of the pressure anisotropy, $p_L/p_T$
for $\tau_0 \bar \mu^{1/3} = 0.2$ and various values of $\bar c$. 
Right: The function $F(x)/x$ for $\tau_0 \bar \mu^{1/3} = 0.2$ and various values of 
$\bar c$. 
In both the figures, the light grey lines are the $3^{rd}$ order hydrodynamic results
(hydro). 
}\label{fig:pressure}
\end{figure}

The behaviour of the pressure anisotropy is shown in 
Fig.~\ref{fig:pressure}, which for our initial conditions is
$p_L/p_T=-\frac{3}{2}$ at $\tau=0$. 
One observes that while the system does not exhibit perfect isotropy
(defined by $p_T=p_L$) for the time extent shown, the pressure
anisotropy matches the (viscous) hydrodynamic result at around $\bar\mu^{1/3}\tau\sim 3$. For all practical purposes, the system may therefore be regarded as 'in-equilibrium'
for all times thereafter.
Conversely, there does not seem to be a unique value of $p_L/p_T$ above which
hydrodynamics is applicable.

We are also able to directly compare our results with those reported in Ref. \cite{Heller:2011ju}. 
Following \cite{Heller:2011ju}, we introduce the quantity 
$\f{F(x)}{x} = \frac{d\ln x}{d \ln \tau}$ for $x \equiv \frac{\tau}{\pi} (-a_4)^{1/4}$.
which is known within $3^{rd}$ order hydrodynamics
(\ref{equ:a4hydro})
\beq
\f{F_{hydro}(x)}{x} =\f{2}{3} + \f{1}{9 \pi x} + \f{1 - \log 2}{27\pi^2 x^2} + \f{ 15 - 2 \pi^2 - 45 \log 2 + 24 \log^2 2 }{927\pi^3 x^3}.
\eeq 
As shown in Fig. \ref{fig:pressure},  our numerical results track 
the $3^{rd}$-order hydrodynamics solution for $x > 0.65$.
This finding is consistent with the result reported in Ref. \cite{Heller:2011ju} 
for rather different initial conditions.

\subsection{Area scaling and analytic approximations}

\begin{figure}
\begin{center}
\includegraphics[width=0.45\linewidth]{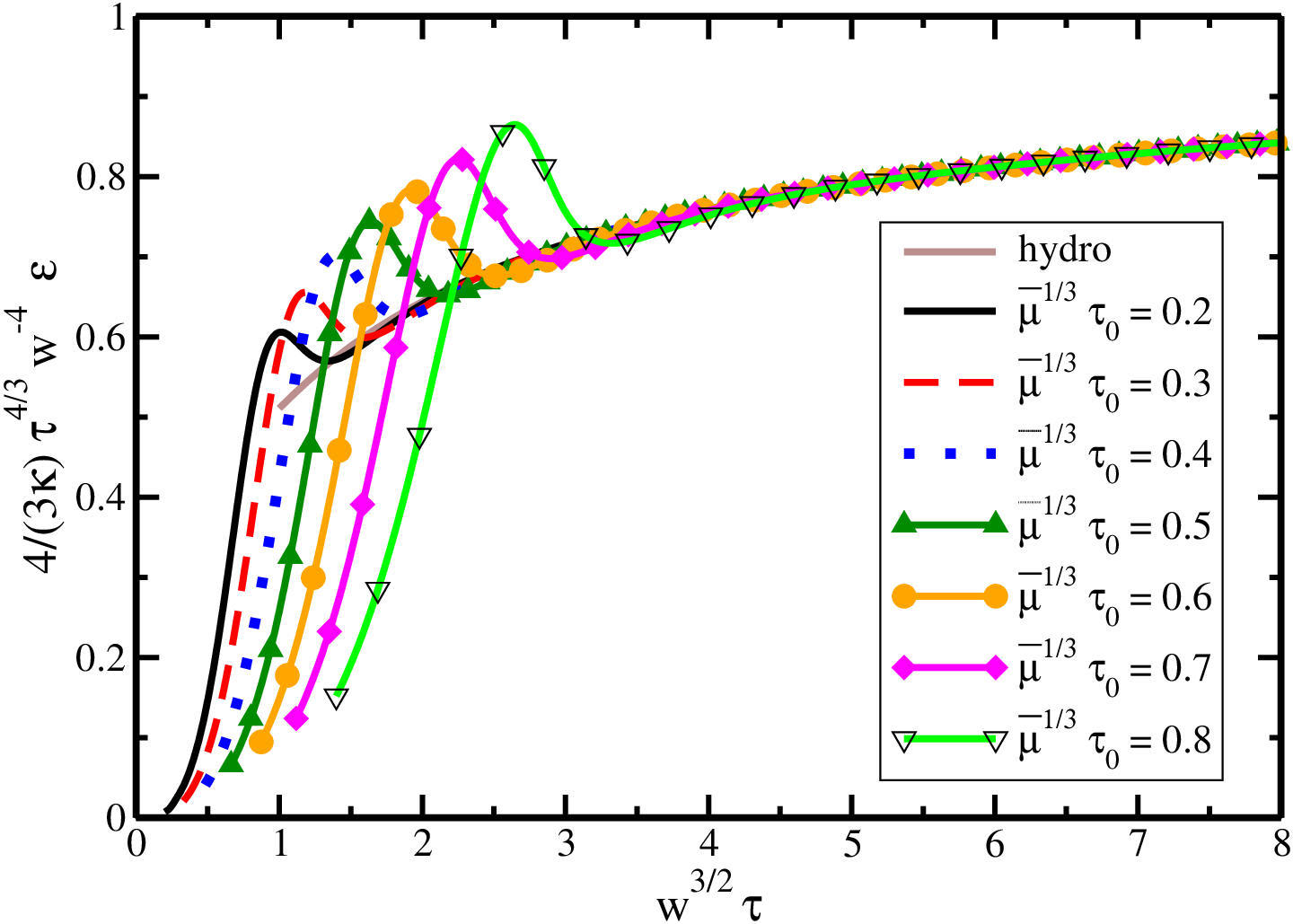}
\hfill
\includegraphics[width=0.45\linewidth]{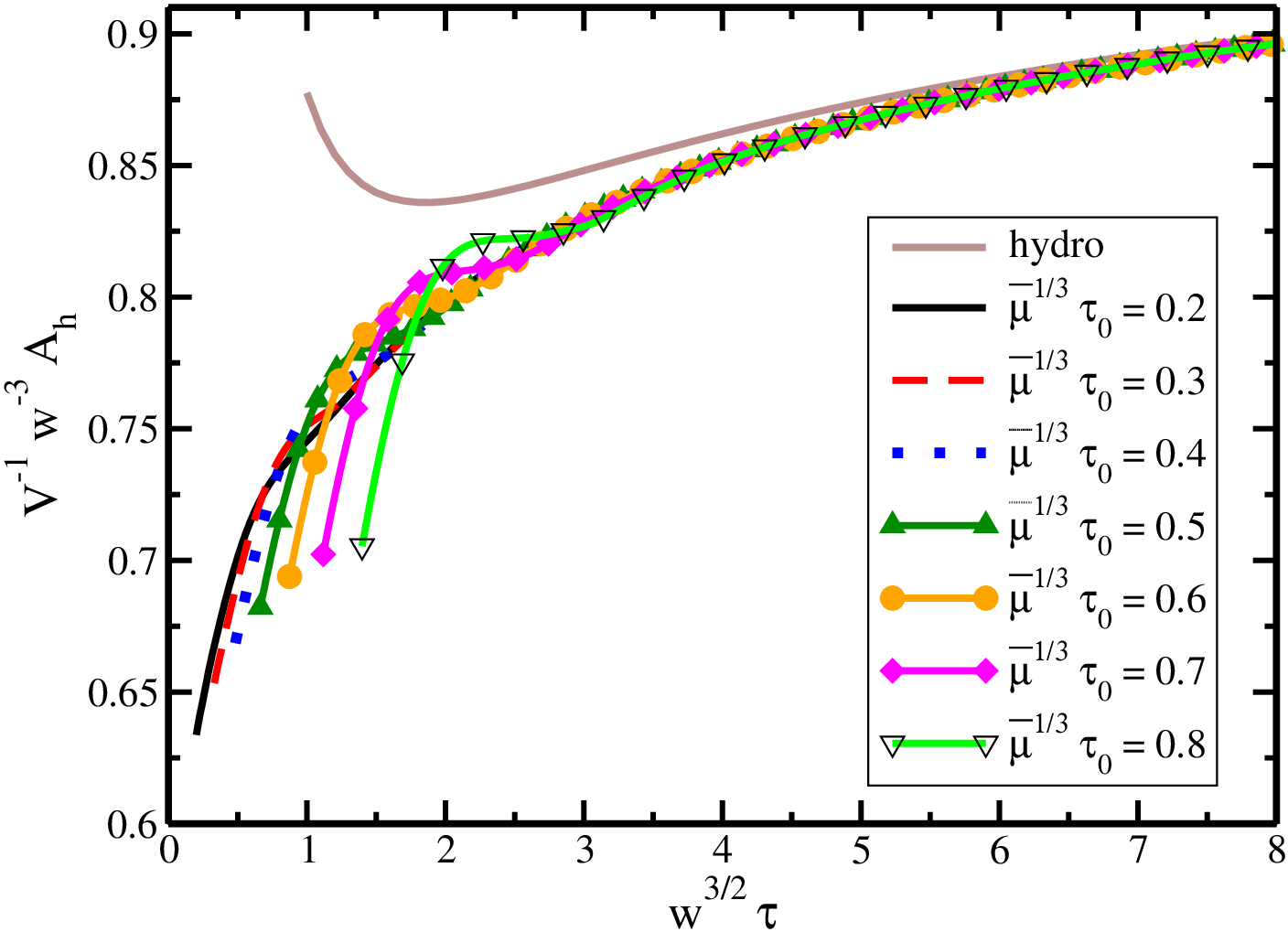}
\end{center}
\caption{The dependence of our numerical results on $\tau_0$ with $\bar c = 16$. Shown
are the energy density (left) and horizon area (right), scaled by the (fitted)
parameter $w(\tau_0,\bar c)$ so that the late time behaviour is universal.
}\label{fig:cbar16}
\end{figure}

While it is numerically hard to send $\tau_0 \bar \mu^{1/3} \rightarrow 0$,
one can hope to learn about the early time behaviour by studying generic values of $\tau_0$.
We thus repeat the above calculations for $\bar c=16$ for different $\tau_0$, always
finding that the late time behaviour is well described by hydrodynamics with 
a parameter $w$ depending on $\tau_0,\bar c$, that is $w(\tau_0,\bar c)$. Extracting 
$w(\tau_0,\bar c)$ by a hydrodynamic fit to the energy density, we may rescale results
for $\epsilon$ and $A_h$ using this quantity so that the late time behaviour becomes
universal. The resulting curves are shown in Fig.~\ref{fig:cbar16}. 

\begin{figure}
\begin{center}
\includegraphics[width=.45\linewidth]{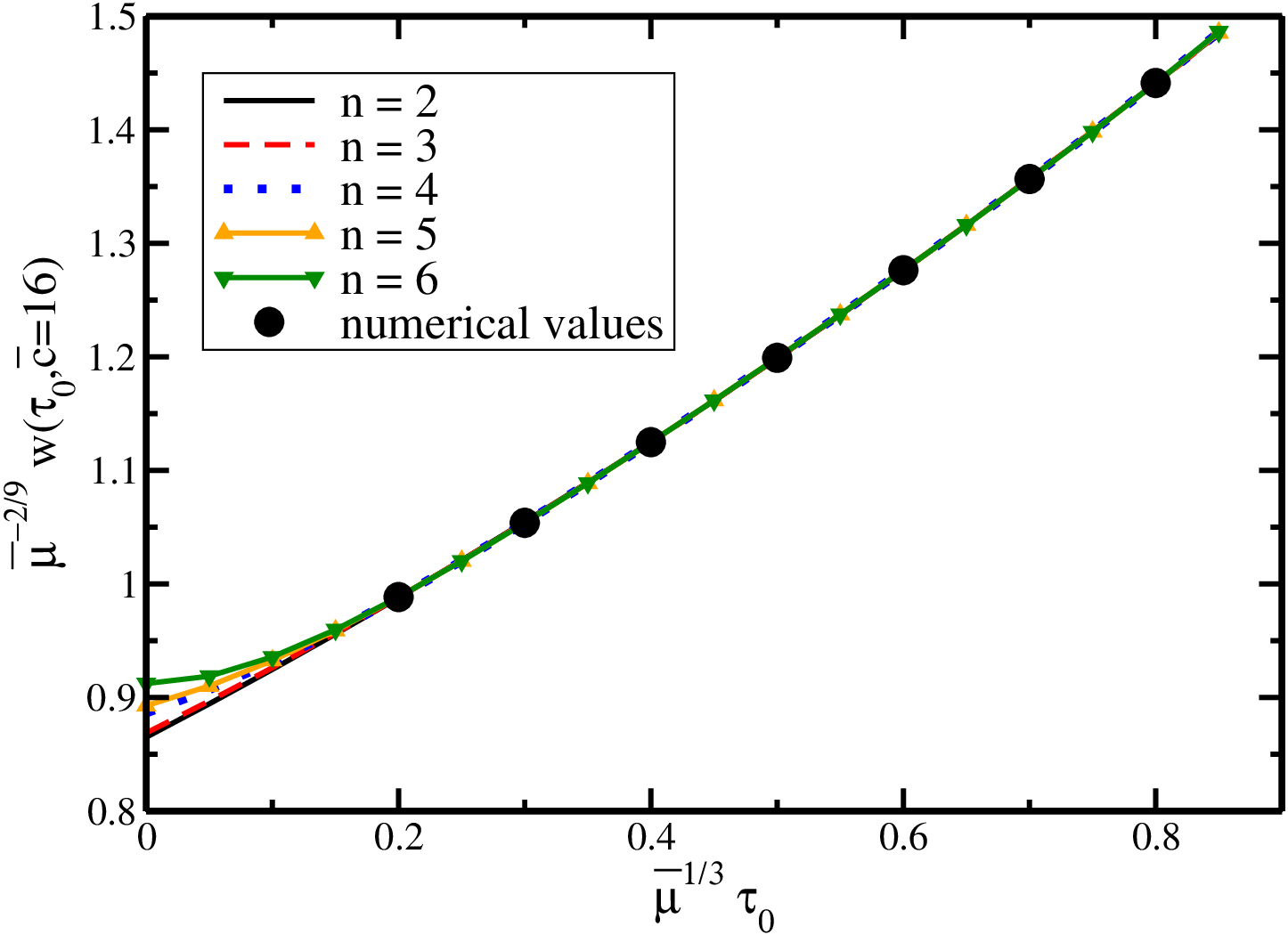}
\hfill
\includegraphics[width=.45\linewidth]{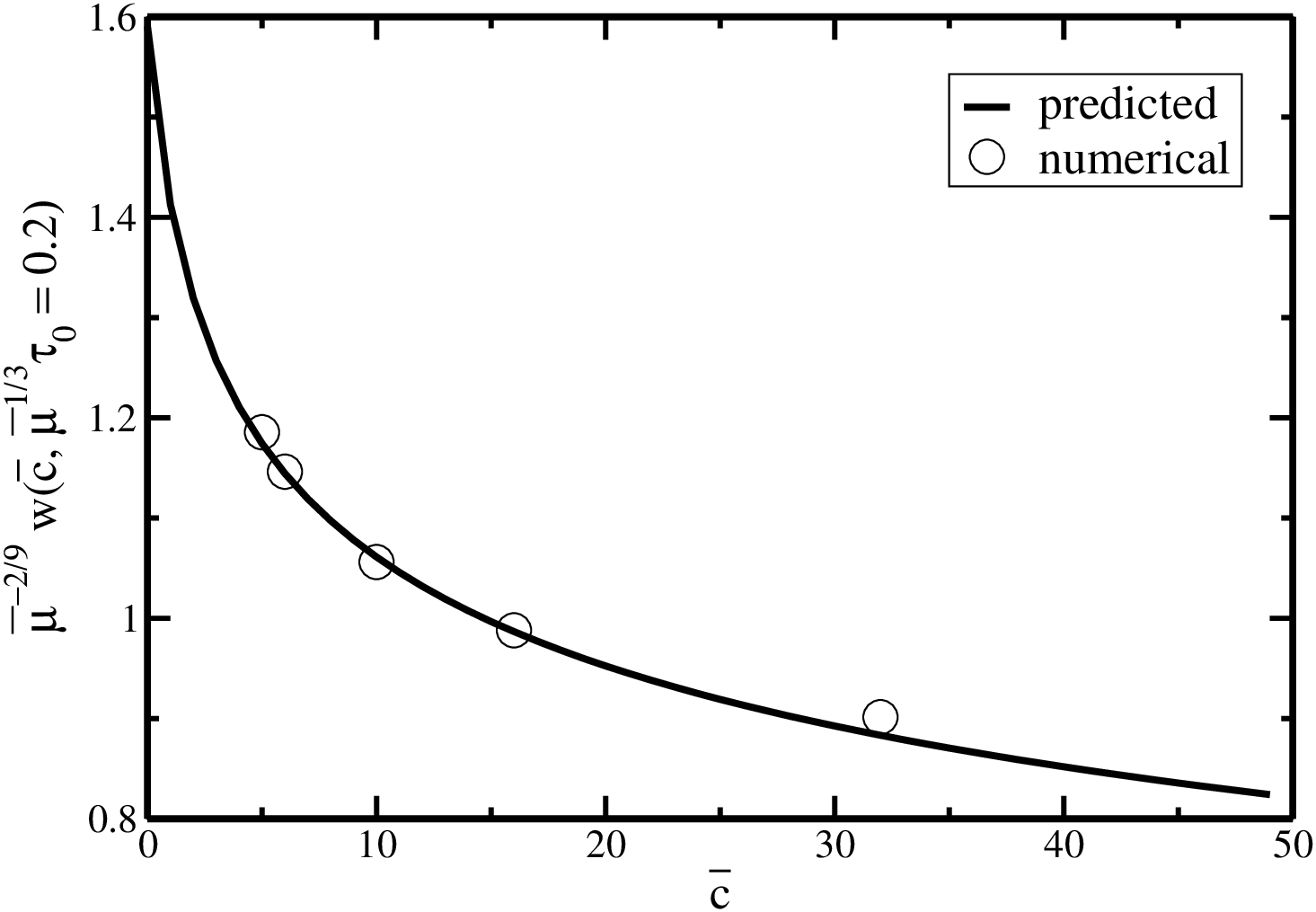}
\end{center}
\caption{Left: fitted numerical values for $w(\tau_0,\bar c)$ as a function of $\tau_0$ for $\bar c = 16$ and polynomial fit of degree $n=2,3,4,5,6$ to the $\tau_0$ dependence.
Right: extracted numerical values of $w(\tau_0,\bar c)$ ('numerical') vs. universal
prediction (\ref{wpred}). 
}\label{fig:wot}
\end{figure}

In Fig. \ref{fig:wot} we plot the extracted values for $w(\tau_0,\bar c)$ as a function
of $\tau_0$. Performing simple polynomial fits with degree $2-6$ we can extrapolate
to $\tau_0\rightarrow 0$, finding the 
value $w(0,\bar c=16)=0.885\pm0.02 \bar\mu^{2/9}$. This suggests that we may try to obtain an analytical
approximation to the time dependence of the horizon area $A_h$ as follows:
since at $\tau_0\rightarrow 0$ the horizon area is given by Eq.~(\ref{PDinitialA}),
we find $V^{-1}A_h(\tau_0=0,\bar c=16)/w(0,\bar c=16)^3\simeq 0.61\pm0.04$. This certainly 
seems consistent with Fig.~\ref{fig:cbar16}. Now knowing the late time behaviour of $A_h$
from hydrodynamics and the initial value from Eq.(\ref{PDinitialA}), we may try to 
interpolate between these two using the ansatz
\beq
\label{Ahansatz}
V^{-1} A_h(\tau,\bar c)/w^3=\frac{u_0+u_1 w(\bar c) \tau^{2/3}}{1+d_1 w(\bar c) \tau^{2/3}}\,
\eeq
where we can fix $u_0,u_1,d_1$ by matching the known late and early time behaviour.
We find $u_1=d_1$, $d_1=\frac{3}{2}(1-u_0)$, $u_0(\bar c=16)=0.61\pm0.04$.
The resulting time dependence is close to the one found in Fig.~\ref{fig:cbar16}, although
it could be further improved by taking into account the known higher order hydrodynamic
coefficients. How does the ansatz (\ref{Ahansatz}) perform for different values of
$\bar c$? To this end, let us simply assume that $u_0=0.61\pm0.04$ for \emph{all}
values of $\bar c$, that is, the behaviour of the horizon area Eq.~(\ref{Ahansatz})
would be a universal function. In this case, it is easy to predict the value 
of $w$ from inverting Eq.~(\ref{Ahansatz}) as 
\beq
\label{wpred}
w(\bar c,\tau_0=0)=\left(\frac{V^{-1}A_h(\tau_0=0,\bar c)}{0.61\pm 0.04}\right)^{1/3}
\eeq
where $A_h(\tau_0=0)$ is given by Eq.~(\ref{PDinitialA}). Since we do not have direct
access to $w(\bar c,\tau_0=0)$, we furthermore assume that for all values of $\bar c$,
the ratio $w(\bar c,\tau_0=0.2)/w(\bar c,\tau_0=0)=\frac{0.988}{0.885\pm0.02}$, i.e.,
the same as for $\bar c=16$. In Fig.~\ref{fig:wot} we then compare the predicted
universal values for $w(\bar c,\bar\mu^{1/3}\tau_0=0.2)$ to the values extracted from our numerical
simulations. Surprisingly, the predictions from the 'pocket formula' (\ref{wpred})
turn out to describe the numerical values almost perfectly! Thus it seems that
--- at least within the class of initial conditions we consider --- the late time
hydrodynamic behaviour is to very good approximation determined by the area of the black
hole horizon at initial times.

\section{A toy model for the early time evolution at RHIC/LHC}
\label{sec:early}

\begin{figure}
\begin{center}
\includegraphics[width=0.48\linewidth]{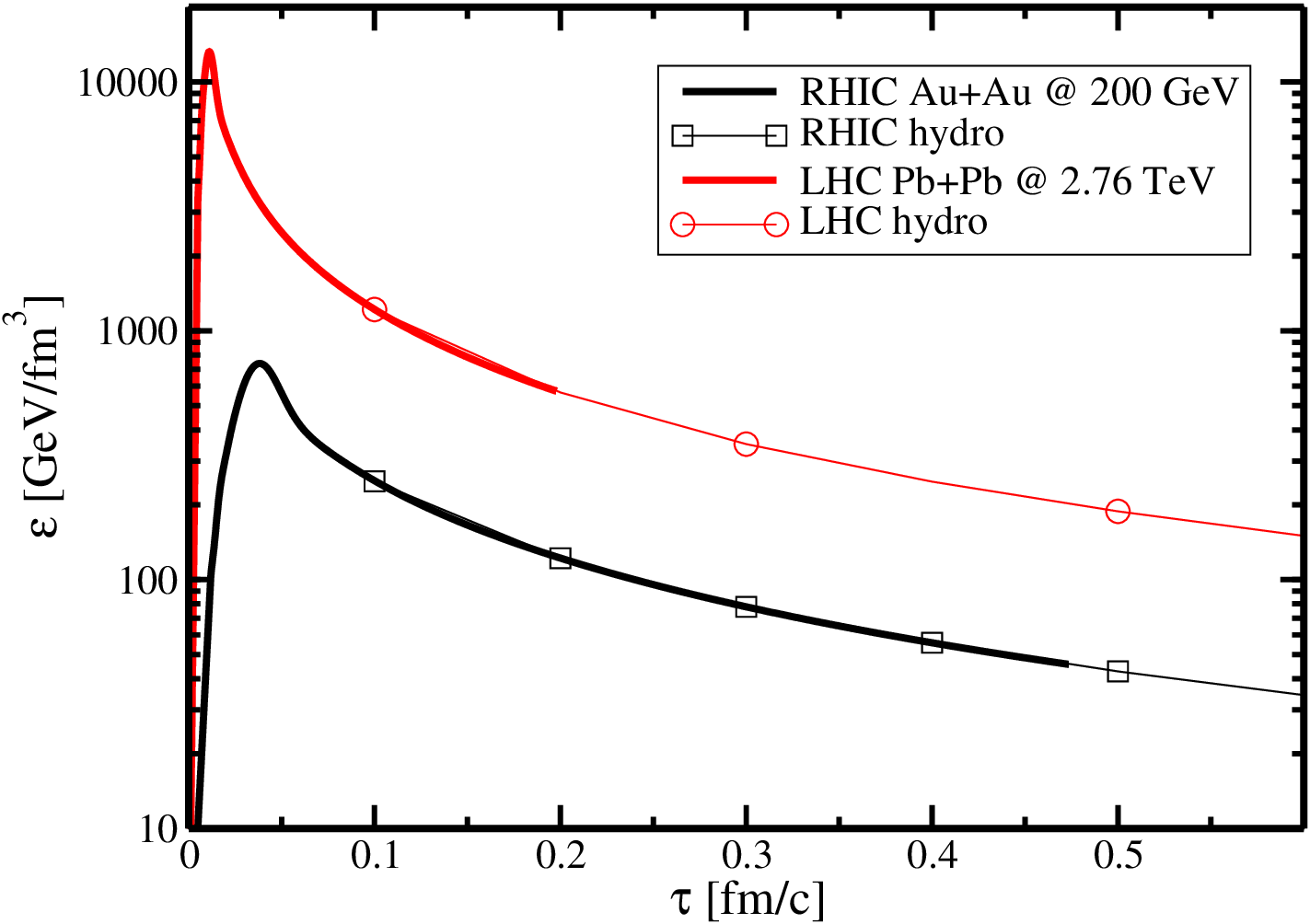}\hfill
\includegraphics[width=0.47\linewidth]{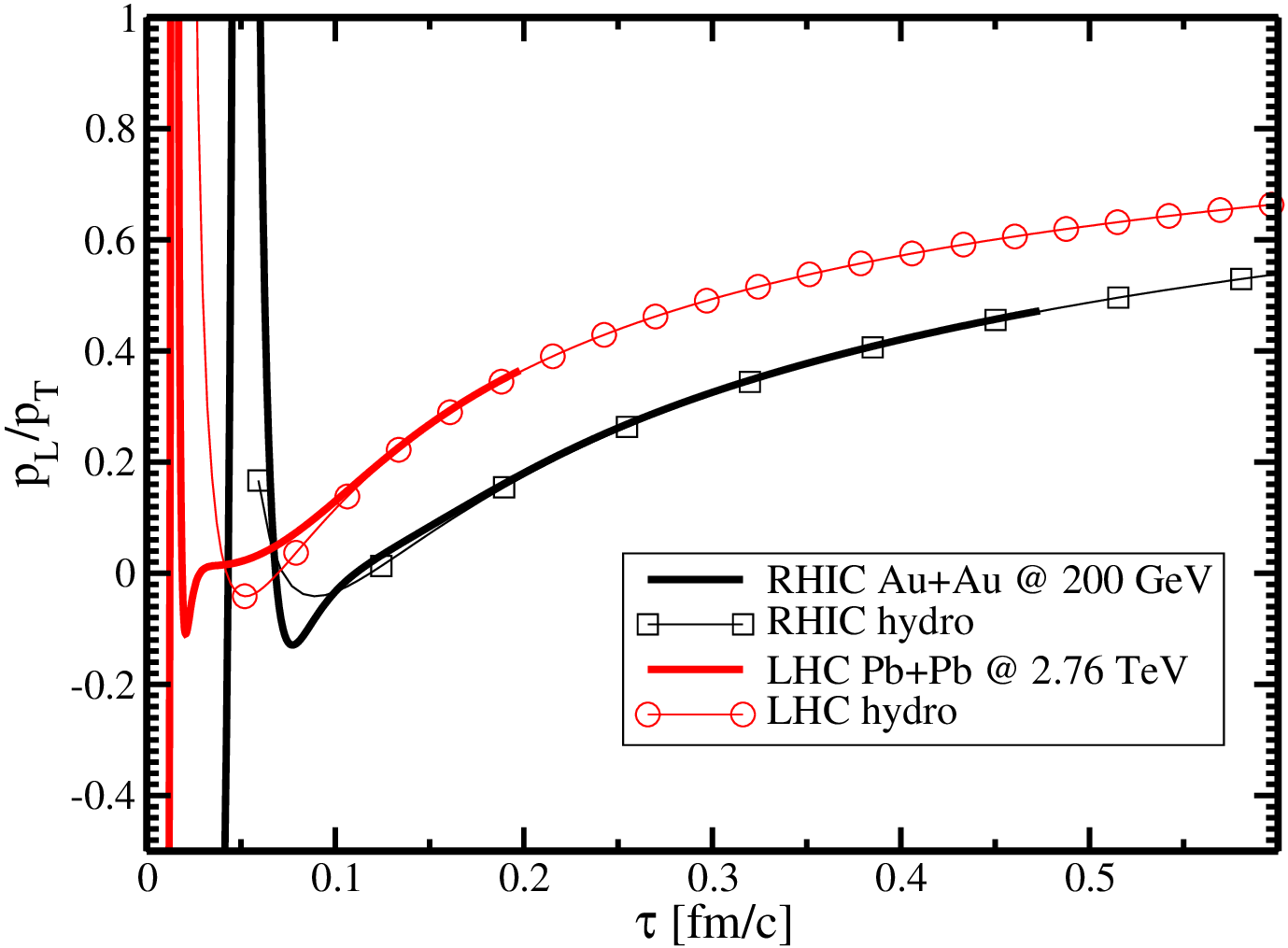}
\end{center}
\caption{Time evolution of the energy density (left) and
pressure anisotropy (right) for RHIC and LHC energies, respectively.
\label{fig:9}
}
\end{figure}

In the preceding sections, we have presented numerical solutions
for the time evolution of energy density and pressure in a strongly
coupled medium that is expanding longitudinally in a boost-invariant
manner. The initial conditions were chosen such as to mimic
those following the collision of two shock waves with transverse
energy density $\mu$ given in Eq.~(\ref{mudef}), where we additionally
introduced a 'fudge parameter' $\bar c$ that (together with $\mu$)
determined the area of the trapped surface at $\tau_0=0$. One may
now ask how well these numerical results correspond to the experimental
situation for heavy-ion collisions encountered at RHIC and the LHC.
Using $a=196,207$ and $R=6.4,6.6$ fm for the atomic number and nuclear radius 
of Au and Pb and $\sqrt{s_{NN}}=200,2760$ GeV for the collision
energies at RHIC and the LHC we have $\mu_{RHIC}\simeq 5.9\ GeV^3$,
$\mu_{LHC}\simeq 81\ GeV^3$. Using our result Eq.~(\ref{wpred}) that
relates the late time behaviour of the trapped surface to that at early times
we can calculate the entropy density 
$s\equiv \frac{\kappa \pi}{\tau V} A_h$
at 'late' times where hydrodynamics applies as
$$
s \simeq \frac{2.84 \pi \mu^{2/3} \kappa^{1/3}}{\tau\sqrt{0.97+\bar c}}\,.
$$
In order to interpret this as QCD entropy density at $\tau=1$ fm/c,
we first need to fix the constant $\kappa$ that is related to 
the number of degrees of freedom we are simulating. At late times
Eqns.~(\ref{equ:a4hydro},\ref{equ:Ahhydro}) 
correspond to $s= \kappa \pi^4 T^3$
with $T=\frac{w}{\pi \tau^{1/3}}$. Since it is known that
this entropy density corresponds to three quarters that of the free 
case, and knowing that 
$s_{QCD}^{\rm free}=\frac{4 (N_c^2-1)+7 N_c N_f}{45} \pi^2 T^3$
we find that we need to set
$$
\kappa = \frac{4 (N_c^2-1)+7 N_c N_f}{60 \pi^2}
$$
or $\kappa\simeq 0.16$ for $N_c=N_f=3$ in order to model QCD.
The corresponding temperatures for $\bar c\rightarrow 0$ are then
$T(\tau=1\ {\rm fm/c})\sim 0.6,1$ GeV for RHIC and LHC energies,
which are much too large. We may reduce the temperatures
to be more in line with values used in actual hydrodynamic
simulations for RHIC and the LHC 
(see e.g. Refs.~\cite{Luzum:2008cw,Luzum:2009sb,Song:2011qa,Niemi:2011ix})
by using the fudge parameter $\bar c$. Additionally, since
we are limited to $\tau_0>0.2 \bar\mu^{1/3}$ one has to rescale
the entropy density by a factor of order one (see preceding section).
In practice, therefore, we choose $\bar c=64$ (RHIC) and $\bar c =512$
(LHC) which with the measured values of $w\bar \mu^{-2/9}=0.8264,0.6587$
give $T(\tau=1\ {\rm fm/c})\sim 0.33,0.41$ GeV for RHIC and LHC,
respectively.
With these parameters, we have a crude, yet  
fully dynamic model for the bulk evolution of the medium created in
heavy-ion collisions from $\tau=0$ to the time when hydrodynamics
becomes applicable. As an example, the evolution of the energy density 
and pressure anisotropy together with the hydrodynamic results
are shown in Fig.~\ref{fig:9}. Note that from the pressure anisotropy 
it seems that hydrodynamics becomes applicable at around $0.15$ fm/c,
regardless of the collision energy. A possible application of our result
would be a calculation of the non-equilibrium photon/dilepton 
production along the lines of 
Refs.~\cite{Martinez:2008di,Bhatt:2010cy,Rebhan:2011ke}
or Upsilon suppression \cite{Strickland:2011mw}, 
which we leave for future work.

\section{Conclusions}

In this work, we have provided numerical solutions to a boost-invariant
(toy) model of shock wave collisions in $AdS_5$, which could be relevant
to the problem of heavy-ion collisions through the AdS/CFT correspondence.
Our initial conditions are such that the initial energy density
evolution is given by the early time analytic 
solution from Ref.~\cite{Grumiller:2008va},
whereas the early time horizon area is controlled by a fudge parameter.
Our numerical results indicate that the late time energy density
behaviour is given by hydrodynamics with a scale parameter that
is determined by the initial black hole horizon area. More work 
is needed to decide whether this is an artefact of the class of initial
conditions we consider or holds true in general. Retuning the number
of degrees of freedom to make our equation of state QCD-like,
and freely choosing the fudge parameter we introduced,
we are able to provide dynamic models for the early time evolution of the
bulk medium following heavy-ion collisions at RHIC and the LHC,
including thermalization of the system. Our results 
may be useful 
for applications such as calculating 
non-equilibrium photon/dilepton production and are available upon request.

\section*{Acknowledgements}

We are indebted to P.~Chesler for providing us with the numerical
code used in Ref.~\cite{cy1} and we would like to thank
P.~Chesler, M.~Martinez, M.~Strickland and Zhe Xu for useful discussions. 
This work was supported in part by the Helmholtz International
Center for FAIR within the framework of the LOEWE program launched
by the state of Hesse.

\appendix
\section{Near-boundary behavior of metric coefficient functions}
\label{app:A}
In this paper, we need to know the following power series expansions of the metric functions near the boundary $r\rightarrow\infty$
\bqa
A_s&=&r^2+a_1 r+\frac{1}{4} \left(a_1^2-4 a_1'\right)+\frac{a_4}{r^2}+\frac{2+3 \tau ^2 a_1^2+\tau ^3 a_1^3+\tau  a_1 \left(4-9 \tau ^4 a_4\right)-6 \tau ^4 (a_4+2 b_4)}{9 \tau ^5 r^3}\nn\\
&+&\frac{1}{60 \tau ^6 r^4}\left(-32+\tau  \left(-37 \tau ^2 a_1^3-10 \tau ^3 a_1^4+6 a_1 \left(-12+10 \tau ^4 (a_4+2 b_4)+3 \tau ^2 a_1'\right)+\tau  a_1^2 \rd\rd\nn\\
&\times&\ld\ld\left(-70+45 \tau ^4 a_4+9 \tau ^2 a_1'\right)+12 \tau  \left(a_1'+\tau ^2 \left(2 a_4+4 b_4-3 \tau  b_4'\right)\right)\right)\right)+\frac{1}{2430 \tau ^7 r^5}\left(2512\rd\nn\\
&+&\ld\tau  \left(1782 \tau ^3 a_1^4+405 \tau ^4 a_1^5+\tau ^2 a_1^3 \left(4289-243 \left(5 \tau ^4 a_4+3 \tau ^2 a_1'\right)\right)-3 \tau  a_1^2 \left(-2162+15 \tau ^2 \rd\rd\rd\nn\\
&\times&\ld\ld\ld\left(54 \tau ^2 (a_4+2 b_4)+43 a_1'-3 \tau  a_1''\right)\right)+2 a_1 \left(2950+9 \tau ^2 \left(-137 a_1'+3 \tau  \left(\tau  \left(-36 a_4-72 b_4\rd\rd\rd\rd\rd\rd\nn\\
&+&\ld\ld\ld\ld\ld\ld 5 a_1'^2+54 \tau  b_4'\right)+5 a_1''\right)\right)\right)-6 \tau  \left(226 a_1'+\tau  \left(-30 a_1''+\tau  \left(80 a_4+160 b_4-45 a_1'^2-138 \tau  b_4' \rd\rd\rd\rd\rd\nn\\
&+&\ld\ld\ld\ld\ld 90 \tau ^2 b_4''\right)\right)\right)\right)\right)\label{equ:As},
\eqa
\bqa
B_s&=&-\frac{2 \log\tau}{3}-\frac{2}{3 \tau  r}+\frac{1+\tau  a_1}{3 \tau ^2 r^2}-\frac{4+3 \tau  a_1 (2+\tau  a_1)}{18 \tau ^3 r^3}+\frac{b_4}{r^4}+\frac{1}{120 \tau ^5 r^5}\left(64+\tau  \left(50 \tau ^2 a_1^3\rd\rd\nn\\
&+&\ld\ld15 \tau ^3 a_1^4+ 10 \tau  a_1^2 \left(10-3 \tau ^2 a_1'\right)-60 a_1 \left(-2+4 \tau ^4 b_4+\tau ^2 a_1'\right)+8 \left(2 \tau ^3 a_4-5 \tau  a_1'\rd\rd\rd\nn\\
&+&\ld\ld\ld 15 \tau ^4 b_4'\right)\right)\right)+\frac{1}{2160 \tau ^6 r^6}\left(-3712+\tau  \left(-1575 \tau ^3 a_1^4-405 \tau ^4 a_1^5+\tau ^2 a_1^3 \left(-4031+1350 \tau ^2 a_1'\right)\rd\rd\nn\\
&+&\ld\ld a_1^2 \left(-7008 \tau +45 \tau ^3 \left(120 \tau ^2 b_4+73 a_1'-7 \tau  a_1''\right)\right)-2 a_1 \left(3742+45 \tau ^2 \left(-47 a_1'\rd\rd\rd\rd\nn\\
&+&\ld\ld\ld\ld\tau  \left(\tau  \left(8 a_4+7 a_1'^2+60 \tau  b_4'\right)+7 a_1''\right)\right)\right)+6 \tau  \left(410 a_1'+\tau  \left(-70 a_1''+\tau \left(-112 a_4\rd\rd\rd\rd\rd\nn\\
&-& \ld\ld\ld\ld\ld 128 b_4+15 \left(-7 a_1'^2+2 \tau  \left(b_4'+7 \tau  b_4''\right)\right)\right)\right)\right)\right)\right),\label{equ:Bs}
\eqa
and
\beq
\Sigma_s=\tau ^{1/3} r+\frac{2+3 \tau  a_1}{6 \tau ^{2/3}}-\frac{1}{9 \tau ^{5/3} r}+\frac{10+9 \tau  a_1}{162 \tau ^{8/3} r^2}+\frac{-40-3 \tau  a_1(20+9 \tau  a_1)}{972 \tau ^{11/3} r^3}+\mathcal{O}\left(\f{1}{r^4}\right),
\eeq
where $a_1$, $a_4$ and $b_4$ are functions only of $\tau$, which satisfy the following equation 
\beq
a_4'= -\frac{2 \left(-2-4 \tau  a_1-3 \tau ^2 a_1^2-\tau ^3 a_1^3+6 \tau ^4 a_4+12 \tau ^4 b_4\right)}{9 \tau ^5}\label{equ:a4}.
\eeq

From Eqns.~(\ref{equ:dotS}) and (\ref{equ:dotB}), one can also get
\bqa
\theta_s &= 
&\frac{\tau  r^4}{4}+\frac{1}{6} (2+3 \tau  a_1) r^3+\frac{1}{8} a_1 (4+3 \tau  a_1) r^2+\frac{1}{8} a_1^2 (2+\tau  a_1) r+\left(\frac{1}{192} a_1^3 (8+3 \tau  a_1)+\frac{1}{4} \tau  a_4\right)\nn\\
&+&\frac{2+\tau  a_1 (4+\tau  a_1 (3+\tau  a_1))-12 \tau ^4 b_4}{30 \tau ^4 r}+\frac{1}{1080 \tau ^5 r^2} \left(-65 \tau ^3 a_1^3-18 \tau ^4 a_1^4+15 \tau ^2 a_1^2 \left(-10+3 \tau ^2 a_1'\right)\right.\nn\\
&+&\left.a_1 \left(-200 \tau +216 \tau ^5 b_4+90 \tau ^3 a_1'\right)-4 \left(28+3 \tau ^2 \left(-5 a_1'+\tau ^2 \left(2 a_4+4 b_4+15 \tau  b_4'\right)\right)\right)\right)\nn\\
&+&\frac{1}{68040 \tau ^6 r^3}\left(10672+\tau  \left(2394 \tau ^3 a_1^4+567 \tau ^4 a_1^5+7 \tau ^2 a_1^3 \left(1049-405 \tau ^2 a_1'\right)\right.\right.\nn\\
&+&\left.\left.3 \tau  a_1^2 \left(5078-3 \tau ^2 \left(785 a_1'+21 \tau  \left(36 \tau  b_4-5 a_1''\right)\right)\right)\right.\right.\nn\\
&+&\left.\left.2 a_1 \left(9490+9 \tau ^2 \left(-575 a_1'+21 \tau  \left(\tau  \left(4 a_4+8 b_4+5 a_1'^2+30 \tau  b_4'\right)+5 a_1''\right)\right)\right)-6 \tau \left(1150 a_1'\right.\right.\right.\nn\\
&+&\left.\left.\left.\tau  \left(-210 a_1''+\tau  \left(-304 a_4-608 b_4-315 a_1'^2+330 \tau  b_4'+630 \tau ^2 b_4''\right)\right)\right)\right)\right)\label{equ:thetas},
\eqa
and
\bqa
\phi_s &=&\frac{r^{3/2}}{3 \sqrt{\tau }}+\frac{(-2+3 \tau  a_1) \sqrt{r}}{12 \tau ^{3/2}}+\frac{(12+\tau  a_1 (-4+3 \tau  a_1)) \sqrt{\frac{1}{r}}}{96 \tau ^{5/2}}\nn\\
&+&\frac{\left(-168-\tau  a_1 (268+5 \tau  a_1 (38+13 \tau  a_1))+768 \tau ^4 b_4\right) \left(\frac{1}{r}\right)^{3/2}}{384 \tau ^{7/2}}\nn\\
&+&\frac{1}{6144 \tau ^{9/2}}\left(5680+\tau  \left(2552 \tau ^2 a_1^3+771 \tau ^3 a_1^4+24 \tau  a_1^2 \left(259-96 \tau ^2 a_1'\right)\right.\right.\nn\\
&-&\left.\left.32 a_1 \left(-287+144 \tau ^2 \left(2 \tau ^2 b_4+a_1'\right)\right)+3072 \left(2 \tau ^3 b_4-\tau  a_1'+3 \tau ^4 b_4'\right)\right)\right) \left(\frac{1}{r}\right)^{5/2}\nn\\
&+&\frac{1}{122880 \tau ^{11/2}}\left(-226144+\tau  \left(-35150 \tau ^3 a_1^4-9615 \tau ^4 a_1^5+8 \tau ^2 a_1^3 \left(-14539+7200 \tau ^2 a_1'\right)\right.\right.\nn\\
&+&\left.\left.16 \tau  a_1^2 \left(-16711+480 \tau ^2 \left(15 \tau ^2 b_4+17 a_1'-3 \tau  a_1''\right)\right)-16 a_1 \left(23083+960 \tau ^2 \right.\right.\right.\nn\\
&\times&\left.\left.\left.\left(-13 a_1'+\tau  \left(10 \tau  b_4+3 \left(\tau  \left(a_1'^2+5 \tau  b_4'\right)+a_1''\right)\right)\right)\right)+1024 \tau \left(140 a_1' \right.\right.\right.\nn\\
&+&\left.\left.\left.\tau  \left(-30 a_1''+\tau  \left(-8 a_4-54 b_4-45 a_1'^2+120 \tau  b_4'+90 \tau ^2 b_4''\right)\right)\right)\right)\right) \left(\frac{1}{r}\right)^{7/2} \label{equ:phis}.
\eqa


\begin{thebibliography}{99}
%
\bibitem{Maldacena}
  J.~M.~Maldacena,
  Adv.\ Theor.\ Math.\ Phys.\  {\bf 2}, 231-252 (1998).
  [hep-th/9711200].
%
\bibitem{Witten}
  E.~Witten,
  Adv.\ Theor.\ Math.\ Phys.\  {\bf 2}, 253-291 (1998).
  [hep-th/9802150].
%
\bibitem{Grumiller:2008va}
  D.~Grumiller, P.~Romatschke,
  JHEP {\bf 0808 } (2008)  027.
  [arXiv:0803.3226].

\bibitem{AlvarezGaume:2008fx}
  L.~Alvarez-Gaume, C.~Gomez, A.~Sabio Vera, A.~Tavanfar, M.~A.~Vazquez-Mozo,
  JHEP {\bf 0902 } (2009)  009.
  [arXiv:0811.3969].


\bibitem{Lin:2009pn}
  S.~Lin, E.~Shuryak,
  Phys.\ Rev.\  {\bf D79 } (2009)  124015.
  [arXiv:0902.1508].


\bibitem{Albacete:2009ji}
  J.~L.~Albacete, Y.~V.~Kovchegov, A.~Taliotis,
  JHEP {\bf 0905 } (2009)  060.
  [arXiv:0902.3046].


\bibitem{Khan:1971vh}
  K.~A.~Khan, R.~Penrose,
  Nature {\bf 229}, 185-186 (1971).
%
\bibitem{cy1}
  P.~M.~Chesler, L.~G.~Yaffe,
  Phys.\ Rev.\ Lett.\  {\bf 102}, 211601 (2009),
  [arXiv:0812.2053]
\bibitem{cy2}
  P.~M.~Chesler, L.~G.~Yaffe,
  Phys.\ Rev.\  {\bf D82}, 026006 (2010).
  [arXiv:0906.4426].
  
\bibitem{Chesler:2010}
  P.~M.~Chesler, L.~G.~Yaffe,
  Phys.\ Rev.\ Lett.\  {\bf 106}, 021601 (2011).
  [arXiv:1011.3562].
%
\bibitem{Heller:2011ju}
  M.~P.~Heller, R.~A.~Janik, P.~Witaszczyk,
  [arXiv:1103.3452].

\bibitem{Bhattacharyya:2009uu}
  S.~Bhattacharyya, S.~Minwalla,
  JHEP {\bf 0909 } (2009)  034.
  [arXiv:0904.0464].


\bibitem{Janik:2005}
  R.~A.~Janik, R.~B.~Peschanski,
  Phys.\ Rev.\  {\bf D73}, 045013 (2006).
  [hep-th/0512162].

\bibitem{Janik:2006}
  R.~A.~Janik, R.~B.~Peschanski,
  Phys.\ Rev.\  {\bf D74}, 046007 (2006).
  [hep-th/0606149].
\bibitem{Kinoshita:2008dq}
  S.~Kinoshita, S.~Mukohyama, S.~Nakamura and K.~y.~Oda,
  Prog.\ Theor.\ Phys.\  {\bf 121}, 121 (2009)
  [arXiv:0807.3797].
  %

\bibitem{Booth:2009ct}
  I.~Booth, M.~P.~Heller, M.~Spalinski,
  Phys.\ Rev.\  {\bf D80}, 126013 (2009).
  [arXiv:0910.0748].
\bibitem{Beuf:2009cx}
  G.~Beuf, M.~P.~Heller, R.~A.~Janik, R.~Peschanski,
  JHEP {\bf 0910 } (2009)  043.
  [arXiv:0906.4423].

\bibitem{Bjorken:1982qr}
  J.~D.~Bjorken,
  Phys.\ Rev.\  {\bf D27}, 140-151 (1983).

\bibitem{Gubser:2008pc}
  S.~S.~Gubser, S.~S.~Pufu, A.~Yarom,
  Phys.\ Rev.\  {\bf D78 } (2008)  066014.
  [arXiv:0805.1551].



\bibitem{Balasubramanian:1999re}
  V.~Balasubramanian, P.~Kraus,
  Commun.\ Math.\ Phys.\  {\bf 208}, 413-428 (1999).
  [hep-th/9902121].
\bibitem{de Haro:2000xn}
  S.~de Haro, S.~N.~Solodukhin, K.~Skenderis,
  Commun.\ Math.\ Phys.\  {\bf 217}, 595-622 (2001).
  [hep-th/0002230].
%
\bibitem{Boyd}
 John~P.~Boyd, 
 ``Chebyshev and Fourier Spectram Methods ($2^{nd}$ Edition),''
New York: Dover (2001) 688 p.


\bibitem{Luzum:2008cw}
  M.~Luzum, P.~Romatschke,
  Phys.\ Rev.\  {\bf C78 } (2008)  034915.
  [arXiv:0804.4015].


\bibitem{Luzum:2009sb}
  M.~Luzum, P.~Romatschke,
  Phys.\ Rev.\ Lett.\  {\bf 103 } (2009)  262302.
  [arXiv:0901.4588].

\bibitem{Song:2011qa}
  H.~Song, S.~A.~Bass, U.~Heinz,
  Phys.\ Rev.\  {\bf C83 } (2011)  054912.
  [arXiv:1103.2380].

\bibitem{Niemi:2011ix}
  H.~Niemi, G.~S.~Denicol, P.~Huovinen, E.~Molnar, D.~H.~Rischke,
  Phys.\ Rev.\ Lett.\  {\bf 106 } (2011)  212302.
  [arXiv:1101.2442].

\bibitem{Strickland:2011mw}
  M.~Strickland,
  %
  [arXiv:1106.2571].

\bibitem{Rebhan:2011ke}
  A.~Rebhan, D.~Steineder,
  %
  [arXiv:1106.3539].

\bibitem{Bhatt:2010cy}
  J.~R.~Bhatt, H.~Mishra, V.~Sreekanth,
  JHEP {\bf 1011 } (2010)  106.
  [arXiv:1011.1969].


\bibitem{Martinez:2008di}
  M.~Martinez, M.~Strickland,
  Phys.\ Rev.\  {\bf C78 } (2008)  034917.
  [arXiv:0805.4552].

\end{thebibliography}
\end{document}